%% file: Main.tex
\documentclass[12pt, draftcls, onecolumn]{IEEEtran}

\IEEEoverridecommandlockouts

\usepackage[cmex10]{amsmath} 
\usepackage{amsfonts,acronym,amssymb,amsthm,dsfont}
\usepackage{graphicx}
\usepackage{url}
\usepackage{cite}
\usepackage{hyperref}
\usepackage{braket}
\usepackage{physics}
\usepackage{enumerate}
\usepackage{lipsum}
\usepackage{diagbox}
\usepackage[dvipsnames]{xcolor}
\usepackage{mathrsfs}
\usepackage{bm}
\usepackage{stackengine}

\makeatletter
\def\blfootnote{\xdef\@thefnmark{}\@footnotetext}
\makeatother

\allowdisplaybreaks
\allowbreak

\input{CommandsAndMacros.tex}
\input{Acronyms.tex}

\allowdisplaybreaks
\begin{document}

\title{Covert Communication Over a Quantum MAC with a Helper}

\author{
\IEEEauthorblockN{Hassan ZivariFard and R\'{e}mi A. Chou and Xiaodong Wang}\\
\thanks{H.~ZivariFard and X.~Wang are with the Department of Electrical Engineering, Columbia University, New York, NY 10027. R.~Chou is with the Department of Computer Science and Engineering, The University of Texas at Arlington, Arlington, TX 76019. The work of H.~ZivariFard and X.~Wang is supported in part by the U.S. Office of Naval Research (ONR) under grant N000142112155 and the work of R.~Chou is supported in part by NSF grant CCF-2047913. E-mails: \{hz2863, xw2008\}@columbia.edu and remi.chou@uta.edu. 
}
}
\maketitle
\date{}
\begin{abstract}
We study covert classical communication over a quantum \ac{MAC} with a helper. Specifically, we consider three transmitters, where one transmitter helps the other two transmitters communicate covertly with a receiver. We demonstrate the feasibility of achieving a positive covert rate over this channel and establish an achievable rate region. Our result recovers as a special case known results for classical communication over classical \acp{MAC} with a degraded message set, classical communication over quantum \acp{MAC}, and classical communication over \acp{MAC} with a helper.  
To the best of our knowledge, our result is the first to achieve covert communication with positive rates over both classical and quantum \acp{MAC}.
\end{abstract}

\section{Introduction}
\label{sec:Intro}
Covert communication ensures that the communication remains undetectable~\cite{LPD_on_AWGN,Reliable_Deniable_Comm,LPD_by_Resolvability,LPD_over_DMC}. In a point-to-point classical \ac{DMC}, it is well established that it is possible to reliably and covertly transmit at most $O(\sqrt{n})$ bits over $n$ channel uses \cite{LPD_by_Resolvability,LPD_over_DMC}, provided that the encoder and decoder share a secret key of size $O(\sqrt{n})$ bits \cite{LPD_by_Resolvability}. Also, covert communication over discrete memoryless \acp{MAC} is studied in \cite{MAC_LPD}, where each transmitter shares a secret key with the receiver. The authors show that each transmitter can transmit on the order of $\sqrt{n}$ reliable and covert bits per $n$ channel use. 
We note that positive covert communication rates can be achieved for point-to-point classical \acp{DMC} \cite{LPD_over_DMC} only when the symbol $x_0\in\calX$, transmitted by the sender in the no-communication mode, is redundant, which implies that the distribution induced on the warden's channel observation by $x_0\in\calX$ can also be induced using the channel input symbols $\{x\in\calX:x\ne x_0\}$. Building on this result, it has been demonstrated that positive covert rates can also be achieved over classical channels under various scenarios: (i) when a friendly jammer is present \cite{UninformedJammer,ISIT22,ISIT21,MyDissertation}, (ii) when the transmitter has access to \ac{CSI} \cite{Covert_With_State,Keyless22,ITW24,Quantum_Covert_MSK}, (iii) when the transmitter has access to \ac{ADSI} \cite{ISIT23,Action_Covert}, (iv) when the warden has uncertainty about the statistical characteristics of its channel \cite{Lee15,Deniable_ITW14}, (v) and when there is a cooperative user who knows either the message or the transmitter's codeword \cite{ISIT22,MyDissertation,Action_Covert}. 

Existing works on covert communication over quantum channels show that optimal rates obey the square root law.  
Specifically, covert communication over bosonic channels is considered in \cite{LPD_on_bosonic,Wang23}, where the authors demonstrate that the covert capacity adheres to the square root law.
Additionally, \cite{Sheikholeslami16,Wang16,Covert_Quantum16} establish that the covert capacity of classical-quantum point-to-point channels also adheres to the square root law. We note that the availability of entangled qubits at both the transmitter and the receiver does not enable a positive covert rate \cite{Entangled_Bosonic}.

In this paper, unlike previous works, we show that it is possible to achieve positive covert rates over a quantum three-user \ac{MAC} when one of the transmitters has access to the private messages of the other two transmitters and aims to assist them in covertly communicating with the receiver. 
We present an achievable covert rate region for this problem. 
The special case of our achievable rate region for the classical and quantum channels recovers as a special case the known results for covert communication over classical \acp{MAC} with degraded message sets \cite{MyDissertation}, for communication over classical \acp{MAC} with a helper \cite{Han_MAC79,ElGamalKim,GunduzSimeone10,Romero17}, and communication over quantum \ac{MAC} \cite{QMAC,Entanglement_MAC}.

Our achievability scheme relies on simultaneous pinching \cite{QIT_Hayashi}, superposition coding, and channel resolvability.  Specifically, our reliability analysis is based on the pinching method, while our covertness analysis employs channel resolvability and pinching techniques \cite{QIT_Hayashi,Hayashi02}, which facilitate the channel resolvability analysis required to establish covertness. In particular, we derive a generalized channel resolvability lemma for quantum \acp{MAC} with a helper.

Some related works that do not consider covertness constraint include: Classical \ac{MAC} with general message sets is studied in \cite{Han_MAC79,GunduzSimeone10,Romero17}. Secure communication over \acp{MAC} is studied in \cite{QMAC_Security}. The problem of communication over \ac{MAC} with cribbing is studied in \cite{WillemsCribbing}, for classical channels, and studied in \cite{UziCribbing}, for quantum channels. Also, secure communication and channel resolvability over classical \acp{MAC} with cribbing is studied in \cite{Helal_Cribbibg} and secure communication over quantum state-dependent channels is studied in \cite{AshnuHayashi2020}.

\textit{Notation:}
Let $\bbN$ be the set of natural numbers and $\bbR$ be the set of real numbers. For any $x, y\in\bbR$, define $\sbra{x}{y} \triangleq [\lfloor x\rfloor, \lceil y \rceil]\cap \bbN$ and $[x] \triangleq \sbra{1}{x}$. For a set of indices $\calI\subset\bbN$, $M_\calI$ denotes $(M_i)_{i\in\calI}$. We denote the set of positive semi-definite operators on a finite-dimensional Hilbert space $\calH$ by $\calP(\calH)$ and denote the set of quantum states by $\calD(\calH)\triangleq\{\rho\in\calP(\calH):\tra[\rho]=1\}$.  
We also denote the space of the bounded linear operators on $\calH$ with $\calL(\calH)$. For $\rho,\sigma\in\calD(\calH)$, the fidelity distance \cite{Jozsa94,Uhlmann76} between these two quantum states is denoted by $\F(\rho,\sigma)\triangleq\lVert\sqrt{\rho}\sqrt{\sigma}\rVert_1$, where $\lVert\rho\rVert_1\triangleq\tra\left[\sqrt{\rho^\dagger\rho}\right]$. For $\rho,\sigma\in\calD(\calH)$, the quantum relative entropy is defined as $\bbD(\rho\lVert\sigma)=\tra\sbr{\rho\pr{\log\rho-\log\sigma}}$. The identity operator on some Hilbert space $\calH$ is denoted by $\bbI$. We use $\tau$ to denote a quantum density matrix and use $\ket{\tau}$ to denote the associated quantum~state. 

\section{Problem Statement}
\label{sec:Problem_Statement}
\begin{figure}[t!]
\centering
\includegraphics[width=8.50cm]{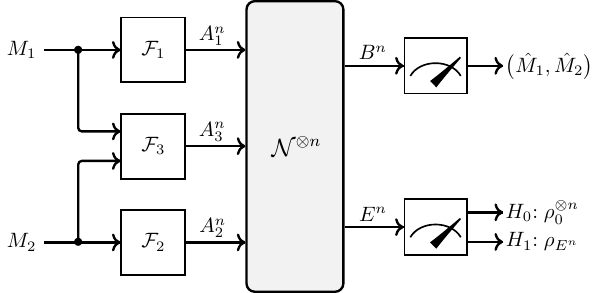}
\caption{Covert communication over a \ac{MAC} with a helper.
}
\label{fig:System_Model}
\vspace{-0.15in}
\end{figure}

\subsection{One-Shot Regime} 
Fig.~\ref{fig:System_Model} illustrates a communication system designed for two transmitters to communicate covertly over a quantum \ac{MAC}. In this setup, a helper, who has access to the private messages of the other transmitters, aims to aid them in covertly communicating with the legitimate receiver. We define codes as follows.
\begin{definition}
\label{defi:code}
A $(2^{R_1},2^{R_2},1)$ code for the quantum channel $\calN_{A_{[3]}\to BE}$ consists of the following:
\begin{itemize}
    \item message sets $\calM_1\triangleq\left[\left\lfloor2^{R_1}\right\rfloor\right]$ and $\calM_2\triangleq\left[\left\lfloor2^{R_2}\right\rfloor\right]$;
    \item an encoding map at the first and the second transmitter, which is a quantum channel  $\calF_{t}$, for $t\in[2]$, that maps the message $M_t\in\calM_t$ to a channel input $A_t\in\calD(\calH)$;
    \item an encoding map at the third transmitter, which is a quantum channel  $\calF_3$, that maps the message $(M_1,M_2)\in\calM_1\times\calM_2$ to a channel input $A_3\in\calD(\calH)$;
    \item a decoding \ac{POVM} $\left\{\calD^{\left(m_{[2]}\right)}_{B\to M_{[2]}}\right\}_{m_{[2]}\in\calM_{[2]}}$, which maps a channel observation $B\in\calD(\calH)$ to $\hat{M}_{[2]}\in\calM_{[2]}$.
\end{itemize}
\end{definition}
The code is known by all the terminals, and the objective is to design a reliable and covert code. 
From Definition~\ref{defi:code} the output of the legitimate receiver's channel is
\begin{subequations}\label{eq:Pe_rho0_CC_CSG}
\begin{align}
    B=\tra_E\calN_{A_{[T]}\to BE}\left(\calF_{[3]\to A_{[3]}}\right),\label{eq:Legi_Output}
\end{align}where $\calF_{[3]\to A_{[3]}}\triangleq\left\{\calF_{t\to A_t}\right\}_{t\in[3]}$. Therefore, from Definition~\ref{defi:code} and \eqref{eq:Legi_Output} the probability of error is defined as
\begin{align}
    P_e&\triangleq\bbP\left\{(M_1,M_2)\ne(\hat{M}_1,\hat{M}_2)\right\}=\frac{1}{\card{\calM_1}\card{\calM_2}}\sum_{(m_1,m_2)\in\calM_1\times\calM_2}\tra\left[\left(\bbI-\calD^{\left(m_{[2]}\right)}_{B\to M_{[2]}}\right)(B)\right].\label{eq:probaility_error}
\end{align}The code $(2^{R_1},2^{R_2},1)$ is reliable if
\begin{align}
    P_e&\le\epsilon.\label{eq:Pe}
\end{align}
\end{subequations}When communication is not happening the transmitters transmit the innocent state $\phi_0^{(t)}\in\calD(\calH_t)$ and therefore the warden receives the quantum state 
\begin{align}
    \rho_0\triangleq\tra_B\calN_{A_{[t]}\to BE}\left(\phi_0^{(1)}\otimes\phi_0^{(2)}\otimes\phi_0^{(3)}\right).\label{eq:No_Comm}
\end{align}

Let $\rho_{M_s}$, for $s\in[2]$, denote the classical-quantum state corresponding to the message $M_s$, which is 
    \begin{align}
        \rho_{M_s}\triangleq\frac{1}{\card{\calM_s}}\sum\limits_{m_s\in\calM_s}\den{m_s}{m_s}.\label{eq:Message}
    \end{align}The state induced at the output of the warden by our code design is denoted by $\tau_E$. Now we define the covertness metric as,
    \begin{align}
        \left\lVert\tau_E- \rho_0\right\rVert_1&\le\delta.\label{eq:Covertness_Metric}
    \end{align}The metric in \eqref{eq:Covertness_Metric} indicates that the state induced at the output of the warden should approximate the state induced at the output of the warden when communication is not happening, i.e., communication is covert.

\begin{definition}[One-Shot]
\label{defi:Code}
    A $S$-pair $(R_1,R_2)$ is said to be achievable for the quantum channel $\calN_{A_{[3]}\to BE}$ if there exists a sequence of $\epsilon$-reliable and $\delta$-covert codes $(2^{R_1},2^{R_2},1)$ that satisfy \eqref{eq:Pe} and \eqref{eq:Covertness_Metric}, respectively. The covert capacity region, denoted by $\calC_{\textup{C}}$, is defined as the supremum of all achievable covert rate tuples.
\end{definition}

\subsection{Asymptotic Regime} 
We now define the problem in the asymptotic regime, where the channel is utilized $n$ times independently as $n$ approaches infinity. The corresponding $\left(2^{nR_1},2^{nR_2}, n\right)$ code for the channel $\calN_{A_{[3]}\to BE}^{\otimes n}$ is similarly defined as Definition~\ref{defi:code}, with the message set $\calM_s\triangleq\left[\left\lfloor{2^{nR_s}}\right\rfloor\right]$, for $s\in[2]$, quantum states $A_{[3]}^n,B^n,E^n\in\calD(\calH^n)$. Then, the output of the legitimate receiver's channel is
\begin{subequations}\label{eq:rho0_Pe_Asym}
\begin{align}
    B^n=\tra_{E^n}\calN_{A_{[3]}^n\to B^nE^n}\left(\calF_{[3]\to A_{[3]}^n}\right),\label{eq:Legi_Output_Asymp}
\end{align}where $\calF_{[3]\to A_{[3]}^n}\triangleq\left\{\calF_{t\to A_t^n}\right\}_{t\in[3]}$.Also, the probability of error is
\begin{align}
    P_e^{(n)}&\triangleq\bbP\left\{(M_1,M_2)\ne(\hat{M}_1,\hat{M}_2)\right\}=\frac{1}{\card{\calM_1}\card{\calM_2}}\sum_{m_{[2]}\in\calM_{[2]}}\tra\left[\left(\bbI-\calD^{\left(m_{[2]}\right)}_{B^n\to M_{[2]}}\right)\left(B^n\right)\right].\label{eq:Pe_Asym}
\end{align}A sequence of codes $\left(2^{nR_1},2^{nR_2}, n\right)$ is reliable if
\begin{align}
    \lim_{n\to\infty}P_e^{(n)}&=0.\label{eq:Pe_Asymp}
\end{align}When communication is absent, the transmitter $t\in[T]$ sends the innocent state $\phi_{t,0}^n \in \mathcal{D}(\mathcal{H}_t^n)$, and consequently, the warden receives the quantum state
\begin{align}
    \rho_0^{\otimes n}\triangleq\tra_{B^n}\calN_{A_{[T]}^n\to B^nE^n}^{\otimes n}\left(\phi_{1,0}^n\otimes\phi_{2,0}^n\otimes\phi_{3,0}^n\right).\label{eq:rho0_Asym}
\end{align}
\end{subequations}
Now we define the covertness metric as,
    \begin{align}
        \lim_{n\to\infty}\left\lVert\tau_{E^n}-\rho_0^{\otimes n}\right\rVert_1&=0.\label{eq:CC_CSK_Metric_Asymp}
    \end{align}
\begin{definition}[Asymptotic]
\label{defi:Asymp_CC}
    A $S$-tuple $(R_1,R_2)$ is said to be achievable for the quantum channel $\calN_{A_{[3]}^n\to B^nE^n}$ if there exists a sequence of codes $\left(2^{nR_1},2^{nR_2},n\right)$ such that \eqref{eq:Pe_Asymp} and \eqref{eq:CC_CSK_Metric_Asymp} hold. The covert capacity region, denoted by $\calC_{\textup{C}}$, is defined as the supremum of all achievable covert rate tuples.
\end{definition}

\section{Main Results}
\subsection{One-Shot Results}
\begin{theorem}[One-shot Achievable Rate Region]
\label{thm:Achievable_One_Shot}
Given a quantum \ac{MAC} with a helper $\calN_{A_{[3]}\to BE}$, and $\rho_{X_{[3]}A_{[3]}}=\sum\limits_{x_{[3]}}p_{X_1}(x_1)p_{X_2}(x_2)p_{X_3\lvert X_1X_2}(x_3\lvert x_1,x_2)\den{x_1}{x_1}_{X_1}\otimes\den{x_2}{x_2}_{X_2}\otimes\den{x_3}{x_3}_{X_3}\otimes\theta_{A_1}^{x_1}\otimes\theta_{A_2}^{x_2}\otimes\theta_{A_3}^{x_3}$, such that $\rho_E=\rho_0$, there exists a $\pr{2^{R_1},2^{R_2},1}$ code such that
\begin{subequations}\label{eq:Pe_Covertness}
\begin{align}
    \bbP\left\{\pr{\hat{M}_1,\hat{M}_2}\ne \pr{M_1,M_2}\right\}&\le6\mu_B^\alpha2^{\alpha\pr{R_1+R_2-\ubar{\D}_{1-\alpha}\pr{\rho_{X_{[3]}B}\lVert\rho_{X_{[3]}}\otimes\rho_B}}}\nonumber\\
    &\quad+\mu_{B,2}^\alpha2^{\alpha\pr{ R_1-\ubar{\I}_{1-\alpha}\pr{X_1,X_3;B\lvert X_2}}}\nonumber\\
    &\quad+\mu_{B,1}^\alpha2^{\alpha\pr{ R_2-\ubar{\I}_{1-\alpha}\pr{X_2,X_3;B\lvert X_1}}},\label{eq:Pe_Thm}\\
    \left\lVert\tau_E- \rho_0\right\rVert_1&\le\frac{2}{\sqrt{\alpha}}\left(\pr{\mu_{E,2}}^{\frac{\alpha}{2}}2^{\frac{\alpha}{2}\pr{-(R_1+R_2)+\ubar{\D}_{1+\alpha}\pr{\rho_{X_{[3]}E}\big\lVert\rho_{X_{[3]}}\otimes\rho_E}}}\right.\nonumber\\
    &\qquad+\pr{\mu_{E,1}}^{\frac{\alpha}{2}}2^{\frac{\alpha}{2}\pr{-R_1+\ubar{\D}_{1+\alpha}\pr{\rho_{X_1E}\big\lVert\rho_{X_1}\otimes\rho_E}}}\nonumber\\
    &\qquad\left.+\pr{\mu_E}^{\frac{\alpha}{2}}2^{\frac{\alpha}{2}\pr{-R_2+\ubar{\D}_{1+\alpha}\pr{\rho_{X_2E}\big\lVert\rho_{X_2}\otimes\rho_E}}}\right),\label{eq:Covertnes_Thm}
    \end{align}where $\alpha\in\left(0,\frac{1}{2}\right)$, and $\mu_{B,2}, \mu_{B,1}, \mu_B, \mu_E, \mu_{E,1}$, and $\mu_{E,2}$ are defined in~\eqref{eq:gs}.
\end{subequations}
\end{theorem}Theorem~\ref{thm:Achievable_One_Shot} is proved in Section~\ref{proof:thm:Achievable_One_Shot}.

\subsection{Asymptotic Results}
\begin{theorem}
\label{thm:Achievable_Asymp}
We define the rate region $\calR_{\textup{q-MAC}}$ as follows,
\begin{align}%
&\calR_{\textup{q-MAC}}\triangleq\bigcup_{\rho_{X_{[3]}A_{[3]}}\in\calJ} %
\left\{ \begin{array}{rl}
  (R_1,R_2):\;
    R_1<I(X_1,X_3;B\lvert X_2),\\
    R_2<I(X_2,X_3;B\lvert X_1),\\
    R_1+R_2<I(X_1,X_2,X_3;B),
	\end{array}
\right\},
\label{eq:inRnone}
\end{align}where
\begin{align}
  \calJ \triangleq \left.\begin{cases}\rho_{X_{[3]}A_{[3]}}:\\
  \rho_{X_{[3]}A_{[3]}}=\sum\limits_{x_{[3]}}p_{X_1}(x_1)p_{X_2}(x_2)p_{X_3\lvert X_1X_2}(x_3\lvert x_1,x_2)\\
  \times\den{x_1}{x_1}_{X_1}\otimes\den{x_2}{x_2}_{X_2}\otimes\den{x_3}{x_3}_{X_3}\otimes\theta_{A_1}^{x_1}\otimes\theta_{A_2}^{x_2}\otimes\theta_{A_3}^{x_3},\\
I(X_1,X_3;B\lvert X_2)>I(X_1,X_3;E\lvert X_2),\\
I(X_2,X_3;B\lvert X_1)>I(X_2,X_3;E\lvert X_1),\\
I(X_1,X_2,X_3;B)>I(X_1,X_2,X_3;E),\\
\rho_E=\rho_0,
\end{cases}\hspace{-2mm}\right\}.\label{eq:thm_S}
\end{align} 
An inner bound for the covert capacity region of a quantum \ac{MAC} $\calN_{A_{[3]}\to BE}$, depicted in Fig.~\ref{fig:System_Model},~is
    \begin{align}
        \calC_{\textup{q-MAC}}\supseteq\calR_{\textup{q-MAC}}.\nonumber
    \end{align}
\end{theorem}Theorem~\ref{thm:Achievable_Asymp} is proved in Section~\ref{proof:thm:Achievable_Asymp}.
\begin{corollary}[Classical Communication Over a Quantum \ac{MAC}]
    By setting  $X_3=A_3=\emptyset$ and removing the covertness constraint $\rho_E=\rho_0$--and thus eliminating the mutual information constraints in \eqref{eq:thm_S}--the achievable rate region described in Theorem~\ref{thm:Achievable_Asymp} reduces to the achievable rate region for classical communication over \acp{MAC}, presented in \cite[Theorem~9]{QMAC}.
\end{corollary}
\begin{corollary}[Communication Over a classical \ac{MAC} with a Helper]
    By removing the covertness constraint $\rho_E=\rho_0$--and thus eliminating the mutual information constraints in \eqref{eq:thm_S}--the achievable rate region described in Theorem~\ref{thm:Achievable_Asymp} reduces to the capacity region for communication over a \ac{MAC} with a helper, presented in \cite[Problem~5.20]{ElGamalKim}, see also \cite{Han_MAC79}.
\end{corollary}
\begin{corollary}[Covert Communication Over a classical \ac{MAC} with one message]
    By setting  $X_2=A_2=\emptyset$ and $M_2=\emptyset$, the achievable rate region described in Theorem~\ref{thm:Achievable_Asymp} reduces to the capacity region for covert communication over the classical \ac{MAC} with one message, presented in \cite[Theorem~36]{MyDissertation}, when there is no secret-shared key between the transmitter and the receiver.
\end{corollary}

\section{Proof of Theorem~\ref{thm:Achievable_One_Shot}}
\label{proof:thm:Achievable_One_Shot} 
We show that for each $\delta_1,\delta_2,\epsilon>0$, there exists a $\pr{1,\epsilon,2^{R_1-\delta_1},2^{R_2-\delta_2}}$ code for the quantum channel $\calN_{A_{[3]}\to BE}$ that satisfies both the reliability constraint and the covertness constraint. 
Fix $p_{X_1},p_{X_2},p_{X_3\lvert X_1X_2}$ and $\delta_1,\delta_2,\epsilon>0$.
\subsection{Random Codebook Generation}
\label{sec:Codebook_Cons} 
Let  $C_1\triangleq\br{X_1(m_1)}_{m_1\in\calM_1}$, where $\calM_1\triangleq\brk{2^{R_1}}$, be a random codebook generated \ac{iid} according to $p_{X_1}$. A realization of $C_1$ is denoted by  $\calC_1\triangleq\br{x_1(m_1)}_{m_1\in\calM_1}$.

Similarly, let  $C_2\triangleq\br{X_2(m_2)}_{m_2\in\calM_2}$, where $\calM_2\triangleq\brk{2^{R_2}}$, be a random codebook generated \ac{iid} according to $p_{X_2}$. A realization of $C_2$ is denoted by  $\calC_2\triangleq\br{x_2(m_2)}_{m_2\in\calM_2}$.

Also, for every $\calC_1$, $\calC_2$, $x_1\in\calC_1$, and $x_2\in\calC_2$, let $C_3\triangleq\big(X_3(m_1,m_2)\big)_{(m_1,m_2)\in\calM_1\times\calM_2}$, be a set of random codeword generated \ac{iid} according to $p_{X_3\lvert X_1X_2}(x_3\lvert x_1(m_1),x_2(m_2))$. A realization of $C_3$ is denoted by $\calC_3\triangleq\big(x_3(m_1,m_2)\big)_{(m_1,m_2)\in\calM_1\times\calM_2}$.
Also, let $C\triangleq\left(C_1,C_2,C_3\right)$ and $\calC\triangleq\left(\calC_1,\calC_2,\calC_3\right)$.

\subsection{Encoding}To send the message $m_t$, for $t\in[2]$, the Transmitter $t$ selects $x_t(m_t)$, and prepares $\rho_{A_t}=\theta^{x_t}$ and transmits it over the channel. Also, given $(m_1,m_2)$, the third transmitter computes $x_3(m_1,m_2)$ and prepares $\rho_{A_3}=\theta^{x_3}$ and transmits it over the channel. This encoding scheme induces the following state at the output of the warden
\begin{align}
    &\tau_{E\lvert\calC}\triangleq\frac{1}{2^{R_1+R_2}}\sum_{m_{[3]}}\rho_{E\lvert x_1\pr{m_1},x_2\pr{m_2},x_3\pr{m_1,m_2}}.\label{eq:Joint_Dist}
\end{align}
\subsection{Pinching}
\label{sec:Pinching}
Our decoding procedure is based on the pinching method \cite{QIT_Hayashi}. Consider states $\sigma\in\calD(\calH)$ and $\rho\in\calD(\calH)$, and let $\sigma=\sum_{i}\lambda_i\den{z_i}{z_i}$ be the spectral decomposition of the state $\sigma$. The pinching operation of the state $\rho$ \ac{wrt} the spectral decomposition of the state $\sigma$ is defined as $\Delta_\sigma(\rho)\triangleq\sum_i\den{z_i}{z_i}\rho\den{z_i}{z_i}=\sum_i\bra{z_i}\rho\ket{z_i}\den{z_i}{z_i}$. Note that the state $\sigma$ and the state $\Delta_\sigma(\rho)$ commute. Now consider the following classical-quantum states,
\begin{subequations}
\begin{align}
    &\rho_{X_{[3]}B}\triangleq\sum_{x_{[3]}}p_{X_1}(x_1)p_{X_2}(x_2)p_{X_3\lvert X_1X_2}(x_3\lvert x_1x_2)\den{x_1}{x_1}_{X_1}\otimes\den{x_2}{x_2}_{X_2}\otimes\den{x_3}{x_3}_{X_3}\otimes\rho_{B\lvert x_{[3]}},\label{eq:general_joint_State}\\
    &\rho_{(X_2,X_3)-X_1-B}\triangleq\sum_{x_{[3]}}p_{X_1}(x_1)\den{x_1}{x_1}_{X_1}\otimes\rho_{X_2 X_3\lvert x_1}\otimes\rho_{B\lvert x_1},\label{eq:PDSG_joint_State2}\\
    &\rho_{(X_1,X_3)-X_2-B}\triangleq\sum_{x_{[3]}}p_{X_2}(x_2)\den{x_2}{x_2}_{X_2}\otimes\rho_{X_1 X_3\lvert x_2}\otimes\rho_{B\lvert x_2},\label{eq:PDSG_joint_State3}
\end{align}where $\rho_{X_1 X_3\lvert x_2}$, $\rho_{B\lvert x_2}$, $\rho_{X_2 X_3\lvert x_1}$, and $\rho_{B\lvert x_1}$ represent the appropriate marginals of the state $\rho_{X_{[3]}B}$ given in \eqref{eq:general_joint_State}. 
\end{subequations}
Let $\Delta_B$ be the pinching operation \ac{wrt} the spectral decomposition of the state $\rho_B$. 
For an arbitrary state $\rho\in\calD(\calH)$, we define 
\begin{align*}
    \Delta_1(\rho)&\triangleq\sum_{x_1}\den{x_1}{x_1}\otimes\Delta_{B|x_1}\pr{\bra{x_1}\rho\ket{x_1}},\\
    \Delta_2(\rho)&\triangleq\sum_{x_2}\den{x_2}{x_2}\otimes\Delta_{B|x_2}\pr{\bra{x_2}\rho\ket{x_2}},
\end{align*}where $\Delta_{B|x_1}$ and $\Delta_{B|x_2}$ are the pinching maps \ac{wrt} the spectral decomposition of $\Delta_B\pr{\rho_{B|x_1}}$ and $\Delta_B\pr{\rho_{B|x_2}}$, respectively.
\begin{subequations}\label{eq:gs}
    \begin{align}
        \mu_B&\triangleq\text{number of distinct eigenvalues of } \rho_B,\label{eq:g1}\\
        \mu_{B,1}&\triangleq\text{maximum number of distinct eigenvalues of } \left\{\Delta_B\pr{\rho_{B\lvert x_1}}\right\}_{x_1},\label{eq:g2}\\
        \mu_{B,2}&\triangleq\text{maximum number of distinct eigenvalues of } \left\{\Delta_B\pr{\rho_{B\lvert x_2}}\right\}_{x_2},\label{eq:g3}
    \end{align}where the maximizations in the above definitions are over $x_1$ and $x_2$, respectively.

We define $\Delta_E, \Delta_{E \mid x_1}$, and $\Delta_{E \mid x_1,x_2}$ as the pinching maps with respect to the spectral decompositions of the states $\rho_E,\Delta_E\bigl(\rho_{E \mid x_1}\bigr)$, and $\Delta_{E \mid x_1}\bigl(\rho_{E \mid x_2}\bigr)$, respectively. We also define the following constants,
    \begin{align}
        \mu_E&\triangleq\text{number of distinct eigenvalues of } \rho_E,\label{eq:g5}\\
        \mu_{E,1}&\triangleq\text{maximum number of distinct eigenvalues of } \left\{\Delta_E\pr{\rho_{E\lvert x_1}}\right\}_{x_1},\label{eq:g6}\\
        \mu_{E,2}&\triangleq\text{maximum number of distinct eigenvalues of } \left\{\Delta_{E\mid X_1}\pr{\rho_{E\lvert x_2}}\right\}_{x_2},\label{eq:g7}
    \end{align}where the maximizations in the above definitions are over $x_1$ and $x_2$, respectively.
\end{subequations}

\subsection{Decoding and Error Probability Analysis}
For any two Hermitian matrices $A$ and $B$, we define the projection $\{A \geq B\}$ as $\sum_{i:\lambda_i \geq 0} P_i$, where $A - B$ has the spectral decomposition $\sum_i \lambda_i P_i$, with $\lambda_i$ being the eigenvalues and $P_i$ being the projection onto the eigenspace corresponding to $\lambda_i$. Now, we define the following projection operators for each set $\calS\subseteq[S]$ of the message indices. 
\begin{subequations}\label{eq:Decoding_Projections}
\begin{align}
    \Pi_{X_{[3]}B}^{(1)}&\triangleq\br{\Delta_B\pr{\rho_{X_{[3]}B}}\ge2^{R_1+R_2}\rho_{X_{[3]}}\otimes\rho_B},\label{eq:Decoding_Projection_1}\\
    \Pi_{X_{[3]}B}^{(2)}&\triangleq\br{\Delta_1\pr{\rho_{X_{[3]}B}}\ge2^{R_2}\Delta_B\pr{\rho_{\pr{X_2,X_3}-X_1-B}}},\label{eq:Decoding_Projection_2}\\
    \Pi_{X_{[3]}B}^{(3)}&\triangleq\br{\Delta_2\pr{\rho_{X_{[3]}B}}\ge2^{R_1}\Delta_B\pr{\rho_{\pr{X_1,X_3}-X_2-B}}}.\label{eq:Decoding_Projection_3}
\end{align}
\end{subequations}  
Also, define $\Pi_{X_{[3]}B} \triangleq \Pi_{X_{[3]}B}^{(1)} \Pi_{X_{[3]}B}^{(2)}\Pi_{X_{[3]}B}^{(3)}$. Note that the projections $\Pi_{X_{[3]}B}^{(1)},\Pi_{X_{[3]}B}^{(2)}$, and $\Pi_{X_{[3]}B}^{(3)}$ are commuting with one another. Moreover, we define the following operator:
\begin{align}
    \Upsilon\pr{m_1,m_2}&\triangleq\tra_{X_{[3]}}\Big[\Pi_{X_{[3]}B}\big(\den{X_1(m_1)}{X_1(m_1)}\otimes\den{X_2(m_2)}{X_2(m_2)}\nonumber\\
    &\qquad\qquad\otimes\den{X_3(m_1,m_2)}{X_3(m_1,m_2)}\otimes\bbI_ B\big)\Big].\label{eq:Gamma_Operator}
\end{align}To obtain a set of \ac{POVM} operators, we normalize \eqref{eq:Gamma_Operator} as follows:
\begin{align}
    \Lambda\pr{m_1,m_2}\triangleq\pr{\sum_{m'_1,m'_2}\Upsilon\pr{m'_1,m'_2}}^{-1}\Upsilon\pr{m_1,m_2}\pr{\sum_{m'_1,m'_2}\Upsilon\pr{m'_1,m'_2}}^{-1}.\label{eq:POVMs}
\end{align}The receiver decodes the messages by applying the \ac{POVM} operators specified in \eqref{eq:POVMs}. The following lemma plays a crucial role in the error analysis.
\begin{lemma}
    \label{lemma:error_analysis}
        For $\alpha\in\pr{0:\frac{1}{2}}$ we have,
    \begin{subequations}\label{eq:Lemma_Error}
    \begin{align}
    \tra\left[\left(\bbI-\Pi_{X_{[3]}B}^{(1)}\right)\rho_{X_{[3]}B}\right]&\le \mu_B^\alpha2^{\alpha(R_1+R_2)}2^{-\alpha\ubar{\D}_{1-\alpha}\pr{\rho_{X_{[3]}B}\lVert\rho_{X_{[3]}}\otimes\rho_B}},\label{eq:Lemma_Error_1}\\
    \tra\left[\left(\bbI-\Pi_{X_{[3]}B}^{(2)}\right)\rho_{X_{[3]}B}\right]&\le \mu_{B,1}^\alpha2^{\alpha R_2}2^{-\alpha\ubar{\I}_{1-\alpha}\pr{X_2,X_3;B\lvert X_1}},\label{eq:Lemma_Error_2}\\
    \tra\left[\left(\bbI-\Pi_{X_{[3]}B}^{(3)}\right)\rho_{X_{[3]}B}\right]&\le \mu_{B,2}^\alpha2^{\alpha R_1}2^{-\alpha\ubar{\I}_{1-\alpha}\pr{X_1,X_3;B\lvert X_2}},\label{eq:Lemma_Error_3}\\
    \tra\left[\Pi_{X_{[3]}B}^{(1)}\left(\rho_{X_{[3]}}\otimes\rho_B\right)\right]&\le \mu_B^\alpha2^{-(1-\alpha)(R_1+R_2)}2^{-\alpha\ubar{\D}_{1-\alpha}\pr{\rho_{X_{[3]}B}\lVert\rho_{X_{[3]}}\otimes\rho_B}},\label{eq:Lemma_Error_4}\\
    \tra\left[\Pi_{X_{[3]}B}^{(2)}\left(\rho_{\pr{X_2,X_3}-X_1-B}\right)\right]&\le \mu_{B,1}^\alpha2^{-(1-\alpha)R_2}2^{-\alpha \ubar{\I}_{1-\alpha}\pr{X_2,X_3;B\lvert X_1}},\label{eq:Lemma_Error_5}\\
    \tra\left[\Pi_{X_{[3]}B}^{(3)}\left(\rho_{\pr{X_1,X_3}-X_2-B}\right)\right]&\le \mu_{B,2}^\alpha2^{-(1-\alpha)R_1}2^{-\alpha\ubar{\I}_{1-\alpha}\pr{X_1,X_3;B\lvert X_2}}.\label{eq:Lemma_Error_6}
    \end{align}
\end{subequations}
\end{lemma}
The proof of Lemma~\ref{lemma:error_analysis} is provided in Section~\ref{proof:lemma_error_analysis}.

\textit{Error Analysis:} 
To bound the probability of error averaged over the random choice of the codebook, it is sufficient to bound $\bbE_C\bbP\br{\pr{\hat{M}_1,\hat{M}_2}\ne\pr{1,1}\big\lvert\pr{M_1,M_2}=\pr{1,1}}$, leveraging the symmetry of the codebook construction. We have,
\begin{align}
    &\bbE_C\bbP\br{\pr{\hat{M}_1,\hat{M}_2}\ne\pr{1,1}\big\lvert\pr{M_1,M_2}=\pr{1,1}}\nonumber\\
    &=\bbE_C\sbr{\tra\sbr{\pr{\sum_{\pr{m'_1,m'_2}\ne\pr{1,1}}\Lambda\pr{m'_1,m'_2}}\rho_{B\lvert X_1(1),X_2(1),X_3(1,1)}}}\nonumber\\
    &=\sum_{\pr{m'_1,m'_2}\ne\pr{1,1}}\bbE_C\sbr{\tra\sbr{\Lambda\pr{m'_1,m'_2}\rho_{B\lvert X_1(1),X_2(1),X_3(1,1)}}}\nonumber\\
    &\le\bbE_C\sbr{\tra\sbr{\pr{\bbI-\Lambda\pr{1,1}}\rho_{B\lvert X_1(1),X_2(1),X_3(1,1)}}}\nonumber\\
    &\le2\bbE_C\sbr{\tra\sbr{\pr{\bbI-\Upsilon\pr{1,1}}\rho_{B\lvert X_1(1),X_2(1),X_3(1,1)}}}\nonumber\\
    &\quad+4\sum_{\pr{m'_1,m'_2}\ne\pr{1,1}}\bbE_C\sbr{\tra\sbr{\Upsilon\pr{m'_1,m'_2}\rho_{B\lvert X_1(1),X_2(1),X_3(1,1)}}},\label{eq:Prob_Err_Analysis}
\end{align}where the last inequality follows from the Hayashi-Nagaoka inequality \cite{Hayashi03}. Now we bound the first term on the \ac{RHS} of \eqref{eq:Prob_Err_Analysis} as follows,
\begin{align}
    &2\bbE_C\sbr{\tra\sbr{\pr{\bbI-\Upsilon\pr{1,1}}\rho_{B\lvert X_1(1)X_2(1)X_3(1,1)}}}\nonumber\\
    &\mathop=\limits^{(a)}2\bbE_C\left[\tra\left[\left(\bbI-\tra_{X_{[3]}}\left[\Pi_{X_{[3]}B}\left(\den{X_1(1)}{X_1(1)}\otimes\den{X_2(1)}{X_2(1)}\otimes\den{X_3(1,1)}{X_3(1,1)}\otimes\bbI_B\right)\right]\right)\right.\right.\nonumber\\
    &\times\rho_{B\lvert X_1(1)X_2(1)X_3(1,1)}\Big]\Big]\nonumber\\
    &\mathop=\limits^{(b)}2\tra\left[\left(\bbI-\Pi_{X_{[3]}B}\right)\rho_{X_1(1)X_2(1)X_3(1,1)B}\right]\nonumber\\
    &=2\tra\left[\left(\bbI-\Pi_{X_{[3]}B}\right)\rho_{X_{[3]}B}\right]\nonumber\\
    &\mathop\le\limits^{(c)}2\tra\left[\left(\bbI-\Pi_{X_{[3]}B}^{(1)}\right)\rho_{X_{[3]}B}\right]+2\tra\left[\left(\bbI-\Pi_{X_{[3]}B}^{(2)}\right)\rho_{X_{[3]}B}\right]+2\tra\left[\left(\bbI-\Pi_{X_{[3]}B}^{(3)}\right)\rho_{X_{[3]}B}\right]\nonumber\\
    &\mathop\le\limits^{(d)}2 \mu_B^\alpha2^{\alpha(R_1+R_2)}2^{-\alpha\ubar{\D}_{1-\alpha}\pr{\rho_{X_{[3]}B}\lVert\rho_{X_{[3]}}\otimes\rho_B}}+2\mu_{B,1}^\alpha2^{\alpha R_2}2^{-\alpha\ubar{\I}_{1-\alpha}\pr{X_2,X_3;B\lvert X_1}}\nonumber\\
    &\quad+2\mu_{B,2}^\alpha2^{\alpha R_1}2^{-\alpha\ubar{\I}_{1-\alpha}\pr{X_1,X_3;B\lvert X_2}},\label{eq:Error_Analysis_1}
\end{align}where
\begin{itemize}
    \item[$(a)$] follows from the definition of $\Upsilon$ in \eqref{eq:Gamma_Operator};
    \item[$(b)$] follows from the linearity of the trace operation and expectation, as well as by taking the expectation \ac{wrt} the random codebook $C$;
    \item[$(c)$] follows from the definition of $\Pi_{X_{[3]}B}$ and since we have
    \begin{subequations}\label{eq:Projection_Inequalities}
    \begin{align}
    0&\preceq\pr{\bbI-\Pi_{X_{[3]}B}^{(1)}}^2=\bbI-\Pi_{X_{[3]}B}^{(1)},\label{eq:Projection_Inequality_1}\\
    0&\preceq\pr{\bbI-\Pi_{X_{[3]}B}^{(2)}\Pi_{X_{[3]}B}^{(3)}}^2=\bbI-\Pi_{X_{[3]}B}^{(2)}\Pi_{X_{[3]}B}^{(3)},\label{eq:Projection_Inequality_2}\\
        0&\preceq\pr{\bbI-\Pi_{X_{[3]}B}^{(1)}-\Pi_{X_{[3]}B}^{(2)}\Pi_{X_{[3]}B}^{(3)}+\Pi_{X_{[3]}B}}^2\nonumber\\
        &=\bbI-\Pi_{X_{[3]}B}^{(1)}-\Pi_{X_{[3]}B}^{(2)}\Pi_{X_{[3]}B}^{(3)}+\Pi_{X_{[3]}B}\nonumber\\
        \Rightarrow&\bbI-\Pi_{X_{[3]}B}\preceq\bbI-\Pi_{X_{[3]}B}^{(1)}+\bbI-\Pi_{X_{[3]}B}^{(2)}\Pi_{X_{[3]}B}^{(3)},\label{eq:Projection_Inequality_3}\\
        0&\preceq\pr{\bbI-\Pi_{X_{[3]}B}^{(2)}-\Pi_{X_{[3]}B}^{(3)}+\Pi_{X_{[3]}B}^{(2)}\Pi_{X_{[3]}B}^{(3)}}^2\nonumber\\
        &=\bbI-\Pi_{X_{[3]}B}^{(2)}-\Pi_{X_{[3]}B}^{(3)}+\Pi_{X_{[3]}B}^{(2)}\Pi_{X_{[3]}B}^{(3)}\nonumber\\
        \Rightarrow&\bbI-\Pi_{X_{[3]}B}^{(2)}\Pi_{X_{[3]}B}^{(3)}\preceq\bbI-\Pi_{X_{[3]}B}^{(1)}+\bbI-\Pi_{X_{[3]}B}^{(3)},\label{eq:Projection_Inequality_4}
    \end{align}consequently, substituting \eqref{eq:Projection_Inequality_4} to the \ac{RHS} of \eqref{eq:Projection_Inequality_3} leads to $\bbI-\Pi_{X_{[3]}B}\preceq\bbI-\Pi_{X_{[3]}B}^{(1)}+\bbI-\Pi_{X_{[3]}B}^{(2)}+\bbI-\Pi_{X_{[3]}B}^{(3)}$;
    \end{subequations}
    \item[$(d)$] follows from Lemma~\ref{lemma:error_analysis}.
\end{itemize}
Now, we bound the second term on the \ac{RHS} of \eqref{eq:Prob_Err_Analysis}. Specifically, we have
\begin{align}
    &4\sum_{\pr{m'_1,m'_2}\ne\pr{1,1}}\bbE_C\sbr{\tra\sbr{\Upsilon\pr{m'_1,m'_2}\rho_{B\lvert X_1(1),X_2(1),X_3(1,1)}}}\nonumber\\
    &=4\sum_{m'_1\ne1}\bbE_C\sbr{\tra\sbr{\Upsilon\pr{m'_1,1}\rho_{B\lvert X_1(1),X_2(1),X_3(1,1)}}}+4\sum_{m'_2\ne1}\bbE_C\sbr{\tra\sbr{\Upsilon\pr{1,m'_2}\rho_{B\lvert X_1(1),X_2(1),X_3(1,1)}}}\nonumber\\
    &\quad+4\sum_{m'_1\ne1}\sum_{m'_2\ne2}\bbE_C\sbr{\tra\sbr{\Upsilon\pr{m'_1,m'_2}\rho_{B\lvert X_1(1),X_2(1),X_3(1,1)}}}.\label{eq:Second_PError}
    \end{align}To bound the first term on the \ac{RHS} of \eqref{eq:Second_PError} we have,
\begin{align}
    &4\sum_{m'_1\ne1}\bbE_C\sbr{\tra\sbr{\Upsilon\pr{m'_1,1}\rho_{B\lvert X_1(1),X_2(1),X_3(1,1)}}}\nonumber\\
    &\mathop=\limits^{(a)}4\sum_{m'_1\ne1}\bbE_C\Big[\tra\Big[\tra_{X_{[3]}}\Big[\Pi_{X_{[3]}B}\big(\den{X_1(m'_1)}{X_1(m'_1)}\otimes\den{X_2(1)}{X_2(1)}\nonumber\\
    &\qquad\qquad\otimes\den{X_3(m'_1,1)}{X_3(m'_1,1)}\otimes\bbI_ B\big)\Big]\rho_{B\lvert X_1(1),X_2(1),X_3(1,1)}\Big]\Big]\nonumber\\
    &\mathop=\limits^{(b)}4\sum_{m'_1\ne1}\tra\Big[\bbE_C\Big[\Pi_{X_{[3]}B}\big(\den{X_1(m'_1)}{X_1(m'_1)}\otimes\den{X_2(1)}{X_2(1)}\nonumber\\
    &\qquad\qquad\otimes\den{X_3(m'_1,1)}{X_3(m'_1,1)}\otimes\bbI_ B\big)\Big]\rho_{B\lvert X_1(1),X_2(1),X_3(1,1)}\Big]\nonumber\\
    &\mathop\le\limits^{(c)}4\times2^{R_1}\tra\sbr{\Pi_{X_{[3]}B}\pr{\rho_{\pr{X_1,X_3}-X_2-B}}}\nonumber\\
    &\mathop\le\limits^{(d)}4\times2^{R_1}\tra\sbr{\Pi_{X_{[3]}B}^{(3)}\pr{\rho_{\pr{X_1,X_3}-X_2-B}}}\nonumber\\
    &\mathop\le\limits^{(e)}4\mu_{B,2}^\alpha2^{\alpha\pr{ R_1-\ubar{\I}_{1-\alpha}\pr{X_1,X_3;B\lvert X_2}}},\label{eq:First_Term_Second_PError}
\end{align}where,
\begin{itemize}
    \item[$(a)$] follows from the definition of $\Upsilon$ in \eqref{eq:Gamma_Operator};
    \item[$(b)$] follows from the linearity of the trace operation and expectation; 
    \item[$(c)$] follows by taking the expectation \ac{wrt} the random codebook $C$, and the symmetry of codebook construction \ac{wrt} the messages.
    \item[$(d)$] follows since $0\preceq\pr{\Pi_{X_{[3]}B}^{(3)}-\Pi_{X_{[3]}B}}^2=\Pi_{X_{[3]}B}^{(3)}-\Pi_{X_{[3]}B}$;
    \item[$(e)$] follows from Lemma~\ref{lemma:error_analysis}.
\end{itemize}Similarly to \eqref{eq:First_Term_Second_PError}, we can bound the second and third terms on the right-hand side of \eqref{eq:Second_PError} and obtain
\begin{subequations}
\begin{align}
    &4\sum_{m'_2\ne1}\bbE_C\sbr{\tra\sbr{\Upsilon\pr{1,m'_2}\rho_{B\lvert X_1(1),X_2(1),X_3(1,1)}}}\nonumber\\
    &\quad\le4\times2^{R_2}\tra\sbr{\Pi_{X_{[3]}B}^{(2)}\pr{\rho_{\pr{X_2,X_3}-X_1-B}}}\nonumber\\
    &\quad\mathop\le\limits^{(a)}4\mu_{B,1}^\alpha2^{\alpha\pr{ R_2-\ubar{\I}_{1-\alpha}\pr{X_2,X_3;B\lvert X_1}}},\label{eq:Second_Term_Second_PError_1}\\
    &4\sum_{(m'_1,m'_2)\ne(1,1)}\bbE_C\sbr{\tra\sbr{\Upsilon\pr{m'_1,m'_2}\rho_{B\lvert X_1(1),X_2(1),X_3(1,1)}}}\nonumber\\
    &\quad\le4\times2^{R_1+R_2}\tra\sbr{\Pi_{X_{[3]}B}^{(1)}\pr{\rho_{X_{[3]}}\otimes\rho_B}}\nonumber\\
    &\quad\mathop\le\limits^{(b)}4\mu_B^\alpha2^{\alpha\pr{R_1+R_2-\ubar{\D}_{1-\alpha}\pr{\rho_{X_{[3]}B}\lVert\rho_{X_{[3]}}\otimes\rho_B}}},\label{eq:Second_Term_Second_PError_2}
\end{align}where $(a)$ and $(b)$ follow from Lemma~\ref{lemma:error_analysis}. 
\end{subequations}
Therefore, considering \eqref{eq:First_Term_Second_PError}, \eqref{eq:Second_Term_Second_PError_1}, and \eqref{eq:Second_Term_Second_PError_2}, the \ac{LHS} of \eqref{eq:Second_PError} is bounded by
\begin{align}
    &4\sum_{\pr{m'_1,m'_2}\ne\pr{1,1}}\bbE_C\sbr{\tra\sbr{\Upsilon\pr{m'_1,m'_2}\rho_{B\lvert X_1(1),X_2(1),X_3(1,1)}}}\nonumber\\
    &\quad\le4\mu_{B,2}^\alpha2^{\alpha\pr{ R_1-\ubar{\I}_{1-\alpha}\pr{X_1,X_3;B\lvert X_2}}}+4\mu_{B,1}^\alpha2^{\alpha\pr{ R_2-\ubar{\I}_{1-\alpha}\pr{X_2,X_3;B\lvert X_1}}}\nonumber\\
    &\quad+4\mu_B^\alpha2^{\alpha\pr{R_1+R_2-\ubar{\D}_{1-\alpha}\pr{\rho_{X_{[3]}B}\lVert\rho_{X_{[3]}}\otimes\rho_B}}}.\label{eq:Error_Analysis_Second_Term}
\end{align}
Now, substituting \eqref{eq:Error_Analysis_1} and \eqref{eq:Error_Analysis_Second_Term} in \eqref{eq:Prob_Err_Analysis} completes the proof of \eqref{eq:Pe_Thm}.

\subsection{Covertness Analysis}
\label{sec:Covertness}
 Let,
    \begin{align}
        \rho_E&\triangleq\sum\limits_{x_{[3]}\in\calX_{[3]}}p_{X_1}(x_1)p_{X_2}(x_2)p_{X_3\lvert X_1X_2}(x_3\lvert x_1,x_2)\rho_{E\lvert x_{[3]}},\label{eq:Lemma_rho}
    \end{align}\sloppy and $\left\{p_{X_1}(x_1)p_{X_2}(x_2)p_{X_3\lvert X_1X_2}(x_3\lvert x_1,x_2),\rho_{x_{[3]}}\right\}_{x_{[3]}\in\calX_{[3]}}$ is a given ensemble. To prove that our code design is also covert, we first bound $\bbE_C\Pu\left(\tau_{E\lvert C},\rho_E\right)$, where $\tau_{E\lvert C}$ is the state induced at the output of the warden by our code design, which is defined in \eqref{eq:Joint_Dist}. Then, we choose the distributions $p_{X_1}(x_1),p_{X_2}(x_2),p_{X_3\lvert X_1X_2}(x_3\lvert x_1,x_2)$, $\theta_{A_1}^{x_1}$, $\theta_{A_2}^{x_2}$, and $\theta_{A_3}^{x_3}$ such that $\rho_E=\rho_0$. 
\begin{lemma}
    \label{lemma:Resolvability}
    Let $\rho_E$ be a classical-quantum state as defined in \eqref{eq:Lemma_rho}. Also, let $C$ be a random codebook as defined in Section~\ref{sec:Codebook_Cons}. Then, 
\begin{align}
    &\bbE_C\left[\ubar{\D}_{1+\alpha}\big(\tau_{E\lvert C}\lVert\rho_E\big)\right]\le\frac{1}{\alpha\ln2}\left(\mu_{E,2}^\alpha2^{\alpha\pr{-(R_1+R_2)+\ubar{\D}_{1+\alpha}\pr{\rho_{X_{[3]}E}\big\lVert\rho_{X_{[3]}}\otimes\rho_E}}}\right.\nonumber\\
    &\qquad\left.+\mu_{E,1}^\alpha2^{\alpha\pr{-R_2+\ubar{\D}_{1+\alpha}\pr{\rho_{X_2E}\big\lVert\rho_{X_2}\otimes\rho_E}}}+\mu_E^\alpha2^{\alpha\pr{-R_1+\ubar{\D}_{1+\alpha}\pr{\rho_{X_1E}\big\lVert\rho_{X_1}\otimes\rho_E}}}\right),\label{eq:res}
\end{align}
where $\tau_{E\lvert C}$ is defined in~\eqref{eq:Joint_Dist}.
\end{lemma}
Lemma~\ref{lemma:Resolvability} is proved in Section~\ref{proof:lemma_Resolvability}.

We have,
\begin{align}
    \bbE_C\Pu\left(\tau_{E\lvert C},\rho_E\right)^2&\mathop\le\limits^{(a)}\bbE_C\left[1-2^{-\ubar{\D}_{1+\alpha}\left(\tau_{E\lvert C}\big\lVert\rho_E\right)}\right]\nonumber\\
    &\mathop\le\limits^{(b)}\ln2\times\bbE_C\left[\ubar{\D}_{1+\alpha}\left(\tau_{E\lvert C}\Big\lVert\rho_E\right)\right]\nonumber\\
    &\mathop\le\limits^{(c)}\frac{1}{\alpha}\left(\mu_{E,2}^\alpha2^{\alpha\pr{-(R_1+R_2)+\ubar{\D}_{1+\alpha}\pr{\rho_{X_{[3]}E}\big\lVert\rho_{X_{[3]}}\otimes\rho_E}}}\right.\nonumber\\
    &\qquad\left.+\mu_{E,1}^\alpha2^{\alpha\pr{-R_2+\ubar{\D}_{1+\alpha}\pr{\rho_{X_2E}\big\lVert\rho_{X_2}\otimes\rho_E}}}+\mu_E^\alpha2^{\alpha\pr{-R_1+\ubar{\D}_{1+\alpha}\pr{\rho_{X_1E}\big\lVert\rho_{X_1}\otimes\rho_E}}}\right),\label{eq:Final_Covertness}
\end{align}where
\begin{itemize}
    \item[$(a)$] follows since for two quantum states $\rho\in\calD(\calH)$ and $\sigma\in\calD(\calH)$ we have \cite[Corollary~4.3]{Tomamichel16},
    \begin{align}
        F^2(\rho,\sigma)=2^{-\ubar{\D}_{1/2}\left(\rho\lVert\sigma\right)}\ge2^{-\ubar{\D}_{1+\alpha}\left(\rho\lVert\sigma\right)};\label{eq:pur_SReny}
    \end{align}
    \item[$(b)$] follows since 
    \begin{align}
        1-2^{-\frac{x}{\ln2}}=1-e^{-x}\le x;\label{eq:Ineq_LN}
    \end{align}
    \item[$(c)$] follows from Lemma~\ref{lemma:Resolvability}.
\end{itemize}Now for two arbitrary states $\rho\in\calD(\calH)$ and $\sigma\in\calD(\calH)$ we have \cite[Theorem~9.3.1]{Wilde_Book}
\begin{align}
    \lVert\rho-\sigma\rVert_1\le2\Pu(\rho,\sigma).\label{eq:Tr_Distance_Purified_Dist}
\end{align}Therefore, considering \eqref{eq:Tr_Distance_Purified_Dist} and the fact that for $a,b\in\bbR^+$ we have $\sqrt{a+b}\le\sqrt{a}+\sqrt{b}$, \eqref{eq:Final_Covertness} completes the proof of \eqref{eq:Covertnes_Thm}.

\section{Proof of Lemma~\ref{lemma:error_analysis}}
\label{proof:lemma_error_analysis}
To prove \eqref{eq:Lemma_Error_1}, we have
\begin{align}
    &\tra\left[\left(\bbI-\Pi_{X_{[3]}B}^{(1)}\right)\rho_{X_{[3]}B}\right]\nonumber\\
    &\quad\mathop=\limits^{(a)}\tra\left[\pr{\Delta_B\left(\bbI-\Pi_{X_{[3]}B}^{(1)}\right)}\rho_{X_{[3]}B}\right]\nonumber\\
    &\quad\mathop=\limits^{(b)}\tra\left[\left(\bbI-\Pi_{X_{[3]}B}^{(1)}\right)\Delta_B\pr{\rho_{X_{[3]}B}}\right]\nonumber\\
    &\quad=\tra\left[\left(\bbI-\Pi_{X_{[3]}B}^{(1)}\right)\pr{\Delta_B\pr{\rho_{X_{[3]}B}}}^{1-\alpha}\pr{\Delta_B\pr{\rho_{X_{[3]}B}}}^\alpha\right]\nonumber\\
    &\quad\mathop\le\limits^{(c)}2^{\alpha(R_1+R_2)}\tra\left[\left(\bbI-\Pi_{X_{[3]}B}^{(1)}\right)\pr{\Delta_B\pr{\rho_{X_{[3]}B}}}^{1-\alpha}\pr{\rho_{X_{[3]}}\otimes\rho_B}^\alpha\right]\nonumber\\
    &\quad\mathop\le\limits^{(d)}2^{\alpha(R_1+R_2)}\tra\left[\pr{\Delta_B\pr{\rho_{X_{[3]}B}}}^{1-\alpha}\pr{\rho_{X_{[3]}}\otimes\rho_B}^\alpha\right]\nonumber\\
    &\quad\mathop=\limits^{(e)}2^{\alpha(R_1+R_2)}\tra\left[\pr{\pr{\rho_{X_{[3]}}\otimes\rho_B}^{\frac{\alpha}{2(1-\alpha)}}\Delta_B\pr{\rho_{X_{[3]}B}}\pr{\rho_{X_{[3]}}\otimes\rho_B}^{\frac{\alpha}{2(1-\alpha)}}}^{1-\alpha}\right]\nonumber\\
    &\quad=2^{\alpha(R_1+R_2)}\tra\left[\pr{\pr{\rho_{X_{[3]}}\otimes\rho_B}^{\frac{\alpha}{2(1-\alpha)}}\Delta_B\pr{\rho_{X_{[3]}B}}\pr{\rho_{X_{[3]}}\otimes\rho_B}^{\frac{\alpha}{2(1-\alpha)}}}\right.\nonumber\\
    &\qquad\times\left.\pr{\pr{\rho_{X_{[3]}}\otimes\rho_B}^{\frac{\alpha}{2(1-\alpha)}}\Delta_B\pr{\rho_{X_{[3]}B}}\pr{\rho_{X_{[3]}}\otimes\rho_B}^{\frac{\alpha}{2(1-\alpha)}}}^{-\alpha}\right]\nonumber\\
    &\quad\mathop=\limits^{(f)}2^{\alpha(R_1+R_2)}\tra\left[\pr{\pr{\rho_{X_{[3]}}\otimes\rho_B}^{\frac{\alpha}{2(1-\alpha)}}\rho_{X_{[3]}B}\pr{\rho_{X_{[3]}}\otimes\rho_B}^{\frac{\alpha}{2(1-\alpha)}}}\right.\nonumber\\
    &\qquad\times\left.\pr{\pr{\rho_{X_{[3]}}\otimes\rho_B}^{\frac{\alpha}{2(1-\alpha)}}\Delta_B\pr{\rho_{X_{[3]}B}}\pr{\rho_{X_{[3]}}\otimes\rho_B}^{\frac{\alpha}{2(1-\alpha)}}}^{-\alpha}\right]\nonumber\\
    &\quad\mathop\le\limits^{(g)}\mu_B^\alpha2^{\alpha(R_1+R_2)}\tra\left[\pr{\pr{\rho_{X_{[3]}}\otimes\rho_B}^{\frac{\alpha}{2(1-\alpha)}}\rho_{X_{[3]}B}\pr{\rho_{X_{[3]}}\otimes\rho_B}^{\frac{\alpha}{2(1-\alpha)}}}\right.\nonumber\\
    &\qquad\times\left.\pr{\pr{\rho_{X_{[3]}}\otimes\rho_B}^{\frac{\alpha}{2(1-\alpha)}}\rho_{X_{[3]}B}\pr{\rho_{X_{[3]}}\otimes\rho_B}^{\frac{\alpha}{2(1-\alpha)}}}^{-\alpha}\right]\nonumber\\
    &\quad=\mu_B^\alpha2^{\alpha(R_1+R_2)}\tra\left[\pr{\pr{\rho_{X_{[3]}}\otimes\rho_B}^{\frac{\alpha}{2(1-\alpha)}}\rho_{X_{[3]}B}\pr{\rho_{X_{[3]}}\otimes\rho_B}^{\frac{\alpha}{2(1-\alpha)}}}^{1-\alpha}\right]\nonumber\\
    &\quad=\mu_B^\alpha2^{\alpha(R_1+R_2)}2^{-\alpha\ubar{\D}_{1-\alpha}\pr{\rho_{X_{[3]}B}\lVert\rho_{X_{[3]}}\otimes\rho_B}},\label{eq:First_Term_Lemma}
\end{align}
    where
\begin{itemize}
    \item[$(a)$] follows since $\Pi^{(1)}_{X_{[3]}B}$ commutes with $\rho_B$;
    \item[$(b)$] follows since for three arbitrary states $\sigma\in\calD(\calH)$, with spectral decomposition $\sigma=\sum_i\lambda_i\den{x_i}{x_i}$, and $\rho_1\in\calD(\calH)$, and $\rho_2\in\calD(\calH)$ we have 
    \begin{align}
    \tra\left[\Delta_\sigma(\rho_1)\rho_2\right]&=\sum_i\bra{x_i}\rho_1\ket{x_i}\tra\left[\den{x_i}{x_i}\rho_2\right]\nonumber\\
    &=\sum_i\bra{x_i}\rho_1\ket{x_i}\tra\left[\bra{x_i}\rho_2\ket{x_i}\right]\nonumber\\
    &=\sum_i\tra\left[\bra{x_i}\rho_1\ket{x_i}\right]\bra{x_i}\rho_2\ket{x_i}\nonumber\\
    &=\sum_i\tra\left[\rho_1\den{x_i}{x_i}\right]\bra{x_i}\rho_2\ket{x_i}\nonumber\\
    &=\sum_i\tra\left[\rho_1\bra{x_i}\rho_2\ket{x_i}\den{x_i}{x_i}\right]\nonumber\\
    &=\tra[\rho_1\Delta_\sigma(\rho_2)];\label{eq:Pinching_Switching}
    \end{align}
    \item[$(c)$] follows from the definition of $\Pi_{X_{[3]}B}^{(1)}$ in \eqref{eq:Decoding_Projection_1} and since $f(x)=x^\beta$, for $\beta\in(0\,,1]$, is a matrix monotone function \cite[Section~1.5]{QIT_Hayashi};
    \item[$(d)$] follows from \eqref{eq:Projection_Inequality_1} and since for $A,B,C\in\calP(\calH)$ and $C\le B$ we have $\tra(CA)\le\tra(BA)$, since $\tra((B-C)A)\ge0$ \cite[Lemma~B.5.2]{RennerDissertation};
    \item[$(e)$] follows since $\Delta_B\pr{\rho_{X_{[T]}B}}$ and $\rho_{X_{[T]}}\otimes\rho_B$ commute;
    \item[$(f)$] follows since for two arbitrary states $\sigma\in\calD(\calH)$, with spectral decomposition $\sigma=\sum_i\lambda_i\den{x_i}{x_i}$, and $\rho\in\calD(\calH)$ we have $\tra\left[\Delta_\sigma(\rho)\sigma\right]=\tra\left[\rho\sigma\right]$ since
        \begin{align}
        \tra[\rho\sigma]&=\sum_i\lambda_i\tra\left[\rho\den{x_i}{x_i}\right]\nonumber\\
        &=\sum_i\lambda_i\tra\left[\bra{x_i}\rho\ket{x_i}\right]\nonumber\\
        &=\tra\left[\Delta_\sigma(\rho)\sigma\right];\label{eq:Pinching_Prop2}
    \end{align}
    \item[$(g)$] follows from the following properties:  
    \begin{enumerate}[i.]
        \item For two arbitrary states $\sigma,\rho\in\calD(\calH)$, we have \cite{Hayashi02}
        \begin{align}
            \rho\preceq \mu\Delta_\sigma(\rho),\label{eq:Pinching_Prop0}
        \end{align}where $\mu$ denotes the number of distinct eigenvalues of $\sigma$.
        \item For $A, B, C \in \mathcal{P}(\mathcal{H})$ with $C \leq B$, it holds that $\operatorname{Tr}(CA) \leq \operatorname{Tr}(BA)$ \cite[Lemma~B.5.2]{RennerDissertation}.  
        \item For any Hermitian matrix $A$ and $B \in \mathcal{P}(\mathcal{H})$, the matrix $ABA$ remains non-negative \cite[Lemma~B.5.1]{RennerDissertation}.  
        \item The function $f(x) = x^{-\beta}$ for $\beta \in (0,1)$ is a matrix anti-monotone function \cite[Chapter~5]{Bhatia_Book}.
        \end{enumerate}
\end{itemize}
Now we prove \eqref{eq:Lemma_Error_2},
\begin{align}
    &\tra\left[\left(\bbI-\Pi_{X_{[3]}B}^{(2)}\right)\rho_{X_{[3]}B}\right]\nonumber\\
    &\quad\mathop=\limits^{(a)}\tra\left[\pr{\Delta_1\left(\bbI-\Pi_{X_{[3]}B}^{(2)}\right)}\rho_{X_{[3]}B}\right]\nonumber\\
    &\quad\mathop=\limits^{(b)}\tra\left[\left(\bbI-\Pi_{X_{[3]}B}^{(2)}\right)\Delta_1\pr{\rho_{X_{[3]}B}}\right]\nonumber\\
    &\quad=\tra\left[\left(\bbI-\Pi_{X_{[3]}B}^{(2)}\right)\pr{\Delta_1\pr{\rho_{X_{[3]}B}}}^{1-\alpha}\pr{\Delta_1\pr{\rho_{X_{[3]}B}}}^\alpha\right]\nonumber\\
    &\quad\mathop\le\limits^{(c)}2^{\alpha R_2}\tra\left[\left(\bbI-\Pi_{X_{[3]}B}^{(2)}\right)\pr{\Delta_1\pr{\rho_{X_{[3]}B}}}^{1-\alpha}\pr{\Delta_B\pr{\rho_{\pr{X_2,X_3}-X_1-B}}}^\alpha\right]\nonumber\\
    &\quad\mathop\le\limits^{(d)}2^{\alpha R_2}\tra\left[\pr{\Delta_1\pr{\rho_{X_{[3]}B}}}^{1-\alpha}\pr{\Delta_B\pr{\rho_{\pr{X_2,X_3}-X_1-B}}}^\alpha\right]\nonumber\\
    &\quad\mathop=\limits^{(e)}2^{\alpha R_2}\tra\left[\pr{\Delta_1\pr{\rho_{X_{[3]}B}}}^{1-\alpha}\pr{\sigma_{X_{[3]}B}}^\alpha\right]\nonumber\\
    &\quad\mathop\le\limits^{(f)}\mu_{B,1}^\alpha2^{\alpha R_2}2^{-\alpha\ubar{\D}_{1-\alpha}\pr{\rho_{X_{[3]}B}\lVert\sigma_{X_{[3]}B}}}\nonumber\\
    &\quad\le \mu_{B,1}^\alpha2^{\alpha R_2}2^{-\alpha\min\limits_{\sigma_{X_{[3]}B}}\ubar{\D}_{1-\alpha}\pr{\rho_{X_{[3]}B}\lVert\sigma_{X_{[3]}B}}}\nonumber\\
    &\quad=\mu_{B,1}^\alpha2^{\alpha R_2}2^{-\alpha \ubar{\I}_{1-\alpha}\pr{X_2,X_3;B\lvert X_1}},\label{eq:Second_Term_Lemma}
\end{align}where
\begin{itemize}
    \item[$(a)$], $(b)$, and $(d)$ follow the same steps as those in \eqref{eq:First_Term_Lemma};
    \item[$(c)$] follows from the definition of $\Pi_{\tilde{X}_{[T]}B}^{(2)}$ in \eqref{eq:Decoding_Projection_2} and since $f(x)=x^\beta$, for $\beta\in(0\,,1]$, is a matrix monotone function \cite[Section~1.5]{QIT_Hayashi};
    \item[$(e)$] follows by defining $\sigma_{X_{[3]}B}\triangleq\Delta_B\pr{\rho_{\pr{X_2,X_3}-X_1-B}}$, note that, assuming the spectral decomposition of $\rho_B$ is  $\rho_B=\sum\limits_{b}\lambda_b\den{b}{b}$, we have
    \begin{align}
    \Delta_B\pr{\rho_{\pr{X_2,X_3}-X_1-B}}&=\sum_{x_1}p_{X_1}\pr{x_1}\den{x_1}{x_1}\otimes\rho_{X_2 X_3\lvert x_1}\otimes\pr{\sum_b(\bra{b}\rho_{B\lvert x_1}\ket{b})\den{b}{b}}\nonumber\\
    &=\sum_{x_{[3]}}\sum_bp_{X_1}\pr{x_1}p_{X_2}\pr{x_2}p_{X_3\lvert X_1X_2}\pr{x_3\lvert x_1,x_2}\pr{\bra{b}\rho_{B\lvert x_1}\ket{b}}\nonumber\\
    &\quad\times\den{x_1}{x_1}\otimes\den{x_2}{x_2}\otimes\den{x_3}{x_3}\otimes\den{b}{b}\nonumber;
    \end{align}
    \item[$(f)$] follows from similar steps to those that follow step $(d)$ in \eqref{eq:First_Term_Lemma}.
\end{itemize}
The proof of \eqref{eq:Lemma_Error_3} is similar to the proof of \eqref{eq:Lemma_Error_2} and is omitted for brevity.

To prove \eqref{eq:Lemma_Error_4} we have,
\begin{align}
    &\tra\left[\Pi_{X_{[3]}B}^{(1)}\left(\rho_{X_{[3]}}\otimes\rho_B\right)\right]\nonumber\\
    &\quad=\tra\left[\Pi_{X_{[3]}B}^{(1)}\left(\rho_{X_{[3]}}\otimes\rho_B\right)^{1-\alpha}\left(\rho_{X_{[3]}}\otimes\rho_B\right)^\alpha\right]\nonumber\\
    &\quad\mathop\le\limits^{(a)}2^{-(1-\alpha)(R_1+R_2)}\tra\left[\Pi_{X_{[3]}B}^{(1)}\pr{\Delta_B\pr{\rho_{X_{[3]}B}}}^{1-\alpha}\left(\rho_{X_{[3]}}\otimes\rho_B\right)^\alpha\right]\nonumber\\
    &\quad\mathop\le\limits^{(b)}2^{-(1-\alpha)(R_1+R_2)}\tra\left[\pr{\Delta_B\pr{\rho_{X_{[3]}B}}}^{1-\alpha}\left(\rho_{X_{[3]}}\otimes\rho_B\right)^\alpha\right]\nonumber\\
    &\quad\mathop=\limits^{(c)}2^{-(1-\alpha)(R_1+R_2)}\tra\left[\pr{\left(\rho_{X_{[3]}}\otimes\rho_B\right)^\frac{\alpha}{2(1-\alpha)}\Delta_B\pr{\rho_{X_{[3]}B}}\left(\rho_{X_{[3]}}\otimes\rho_B\right)^\frac{\alpha}{2(1-\alpha)}}^{1-\alpha}\right]\nonumber\\
    &\quad=2^{-(1-\alpha)(R_1+R_2)}\tra\left[\pr{\left(\rho_{X_{[3]}}\otimes\rho_B\right)^\frac{\alpha}{2(1-\alpha)}\Delta_B\pr{\rho_{X_{[3]}B}}\left(\rho_{X_{[3]}}\otimes\rho_B\right)^\frac{\alpha}{2(1-\alpha)}}\right.\nonumber\\
    &\qquad\times\left.\pr{\left(\rho_{X_{[3]}}\otimes\rho_B\right)^\frac{\alpha}{2(1-\alpha)}\Delta_B\pr{\rho_{X_{[3]}B}}\left(\rho_{X_{[3]}}\otimes\rho_B\right)^\frac{\alpha}{2(1-\alpha)}}^{-\alpha}\right]\nonumber\\
    &\quad\mathop=\limits^{(d)}2^{-(1-\alpha)(R_1+R_2)}\tra\left[\pr{\left(\rho_{X_{[3]}}\otimes\rho_B\right)^\frac{\alpha}{2(1-\alpha)}\rho_{X_{[3]}B}\left(\rho_{X_{[3]}}\otimes\rho_B\right)^\frac{\alpha}{2(1-\alpha)}}\right.\nonumber\\
    &\qquad\times\left.\pr{\left(\rho_{X_{[3]}}\otimes\rho_B\right)^\frac{\alpha}{2(1-\alpha)}\Delta_B\pr{\rho_{X_{[3]}B}}\left(\rho_{X_{[3]}}\otimes\rho_B\right)^\frac{\alpha}{2(1-\alpha)}}^{-\alpha}\right]\nonumber\\
    &\quad\mathop\le\limits^{(e)}\mu_B^\alpha2^{-(1-\alpha)(R_1+R_2)}\tra\left[\pr{\left(\rho_{X_{[3]}}\otimes\rho_B\right)^\frac{\alpha}{2(1-\alpha)}\rho_{X_{[3]}B}\left(\rho_{X_{[3]}}\otimes\rho_B\right)^\frac{\alpha}{2(1-\alpha)}}\right.\nonumber\\
    &\qquad\times\left.\pr{\left(\rho_{X_{[3]}}\otimes\rho_B\right)^\frac{\alpha}{2(1-\alpha)}\rho_{X_{[3]}B}\left(\rho_{X_{[3]}}\otimes\rho_B\right)^\frac{\alpha}{2(1-\alpha)}}^{-\alpha}\right]\nonumber\\
    &\quad=\mu_B^\alpha2^{-(1-\alpha)(R_1+R_2)}\tra\left[\pr{\left(\rho_{X_{[3]}}\otimes\rho_B\right)^\frac{\alpha}{2(1-\alpha)}\rho_{X_{[3]}B}\left(\rho_{X_{[3]}}\otimes\rho_B\right)^\frac{\alpha}{2(1-\alpha)}}^{1-\alpha}\right]\nonumber\\
    &\quad=\mu_B^\alpha2^{-(1-\alpha)(R_1+R_2)}2^{-\alpha\ubar{\D}_{1-\alpha}\pr{\rho_{X_{[3]}B}\lVert\rho_{X_{[3]}}\otimes\rho_B}},\label{eq:Third_Term_Lemma}
\end{align}where
\begin{itemize}
    \item[$(a)$] follows from the definition of $\Pi_{X_{[3]}B}^{(1)}$ in \eqref{eq:Decoding_Projection_1} and since $f(x)=x^\beta$, for $\beta\in(0\,,1]$, is a matrix monotone function \cite[Section~1.5]{QIT_Hayashi};
    \item[$(b)$] follows from \eqref{eq:Projection_Inequality_1} and since for $A,B,C\in\calP(\calH)$ and $C\le B$ we have $\tra(CA)\le\tra(BA)$, since $\tra((B-C)A)\ge0$ \cite[Lemma~B.5.2]{RennerDissertation};
    \item[$(c)$] follows since $\Delta_B\pr{\rho_{X_{[3]}B}}$ and $\rho_{X_{[3]}}\otimes\rho_B$ commute;
    \item[$(d)$] follows from \eqref{eq:Pinching_Prop2};
    \item[$(e)$] follows from a similar argument as those used in the step $(f)$ of \eqref{eq:First_Term_Lemma}.
\end{itemize}
We now prove \eqref{eq:Lemma_Error_5},
\begin{align}
    &\tra\left[\Pi_{X_{[3]}B}^{(2)}\left(\rho_{\pr{X_2,X_3}-X_1-B}\right)\right]\nonumber\\
    &\quad\mathop=\limits^{(a)}\tra\left[\pr{\Delta_B\pr{\Pi_{X_{[3]}B}^{(2)}}}\left(\rho_{\pr{X_2,X_3}-X_1-B}\right)\right]\nonumber\\
    &\quad\mathop=\limits^{(b)}\tra\left[\Pi_{X_{[3]}B}^{(2)}\Delta_B\pr{\rho_{\pr{X_2,X_3}-X_1-B}}\right]\nonumber\\
    &\quad=\tra\left[\Pi_{X_{[3]}B}^{(2)}\pr{\Delta_B\pr{\rho_{\pr{X_2,X_3}-X_1-B}}}^{1-\alpha}\pr{\Delta_B\pr{\rho_{\pr{X_2,X_3}-X_1-B}}}^\alpha\right]\nonumber\\
    &\quad\mathop\le\limits^{(c)}2^{-(1-\alpha)R_2}\tra\left[\Pi_{X_{[3]}B}^{(2)}\pr{\Delta_1\pr{\rho_{X_{[3]}B}}}^{1-\alpha}\pr{\Delta_B\pr{\rho_{\pr{X_2,X_3}-X_1-B}}}^\alpha\right]\nonumber\\
    &\quad\mathop\le\limits^{(d)}2^{-(1-\alpha)R_2}\tra\left[\pr{\Delta_1\pr{\rho_{X_{[3]}B}}}^{1-\alpha}\pr{\Delta_B\pr{\rho_{\pr{X_2,X_3}-X_1-B}}}^\alpha\right]\nonumber\\
    &\quad\mathop=\limits^{(e)}2^{-(1-\alpha)R_2}\tra\left[\pr{\Delta_1\pr{\rho_{X_{[3]}B}}}^{1-\alpha}\pr{\sigma_{X_{[3]}B}}^\alpha\right]\nonumber\\
    &\quad\mathop\le\limits^{(f)}\mu_{B,1}^\alpha2^{-(1-\alpha)R_2}2^{-\alpha\ubar{\D}_{1-\alpha}\pr{\rho_{X_{[3]}B}\lVert\sigma_{X_{[3]}B}}}\nonumber\\
    &\quad\le \mu_{B,1}^\alpha2^{-(1-\alpha)R_2}2^{-\alpha\min\limits_{\sigma_{X_{[3]}B}}\ubar{\D}_{1-\alpha}\pr{\rho_{X_{[3]}B}\lVert\sigma_{X_{[3]}B}}}\nonumber\\
    &\quad=\mu_{B,1}^\alpha2^{-(1-\alpha)R_2}2^{-\alpha \ubar{\I}_{1-\alpha}\pr{X_2,X_3;B\lvert X_1}},\label{eq:Fourth_Term_Lemma}
\end{align}where
\begin{itemize}
    \item[$(a)$] follows follows since $\Pi^{(1)}_{X_{[3]}B}$ commutes with $\rho_B$;
    \item[$(b)$] follows from \eqref{eq:Pinching_Switching};
    \item[$(c)$] follows from \eqref{eq:Decoding_Projection_2} and since $f(x)=x^\beta$, for $\beta\in(0\,,1]$, is a matrix monotone function \cite[Section~1.5]{QIT_Hayashi};
    \item[$(d)$] follows from \eqref{eq:Projection_Inequality_2} and since for $A,B,C\in\calP(\calH)$ and $C\le B$ we have $\tra(CA)\le\tra(BA)$, since $\tra((B-C)A)\ge0$ \cite[Lemma~B.5.2]{RennerDissertation};
    \item[$(e)$] follows from the same reasoning as the argument used in step $(e)$ of \eqref{eq:Second_Term_Lemma};
    \item[$(f)$] follows from similar steps to those that follow step $(d)$ in \eqref{eq:First_Term_Lemma}.
\end{itemize}
The proof of \eqref{eq:Lemma_Error_6} is similar to that of \eqref{eq:Lemma_Error_5} and is omitted for brevity.
\section{Proof of Lemma~\ref{lemma:Resolvability}}
\label{proof:lemma_Resolvability}
From the concavity of the $\log$ function and Jensen's inequality, we have
\begin{align}
    \bbE_C\left[\ubar{\D}_{1+\alpha}\big(\tau_{E\lvert C}\lVert\rho_E\big)\right]&\le\frac{1}{\alpha}\log_2\left(\bbE_C\left[2^{\alpha\ubar{\D}_{1+\alpha}\big(\tau_{E\lvert C}\lVert\rho_E\big)}\right]\right).\label{eq:Jensen_Res}
\end{align}
Now, we bound the argument of the $\log$ function in \eqref{eq:Jensen_Res} as,
\begin{align}
    &\bbE_C\sbr{2^{\alpha\ubar{\D}_{1+\alpha}\big(\tau_{E\lvert C}\lVert\rho_E\big)}}\nonumber\\
    &=\bbE_C\tra\left[\left(\rho_E^{-\frac{\alpha}{2(1+\alpha)}}\tau_{E\lvert C}\rho_E^{-\frac{\alpha}{2(1+\alpha)}}\right)^{1+\alpha}\right]\nonumber\\
    &=\bbE_C\tra\left[\left(\rho_E^{-\frac{\alpha}{2(1+\alpha)}}\frac{1}{2^{(R_1+R_2)}}\sum_{m_1,m_2}\rho_{E\lvert X_1\pr{m_1},X_2\pr{m_2},X_3\pr{m_1,m_2}}\rho_E^{-\frac{\alpha}{2(1+\alpha)}}\right)^{1+\alpha}\right]\nonumber\\
    &=\bbE_C\tra\left[\left(\rho_E^{-\frac{\alpha}{2(1+\alpha)}}\frac{1}{2^{(R_1+R_2)}}\sum_{m_1,m_2}\rho_{E\lvert X_1\pr{m_1},X_2\pr{m_2},X_3\pr{m_1,m_2}}\rho_E^{-\frac{\alpha}{2(1+\alpha)}}\right)\right.\nonumber\\
    &\left.\times\left(\rho_E^{-\frac{\alpha}{2(1+\alpha)}}\frac{1}{2^{(R_1+R_2)}}\sum_{m_{[3]}}\rho_{E\lvert X_1\pr{m_1},X_2\pr{m_2},X_3\pr{m_1,m_2}}\rho_E^{-\frac{\alpha}{2(1+\alpha)}}\right)^\alpha\right]\nonumber\\
    &\mathop=\limits^{(a)}\frac{1}{2^{(R_1+R_2)}}\sum_{m_1,m_2}\bbE_C\tra\left[\left(\rho_E^{-\frac{\alpha}{2(1+\alpha)}}\rho_{E\lvert X_1\pr{m_1},X_2\pr{m_2},X_3\pr{m_1,m_2}}\rho_E^{-\frac{\alpha}{2(1+\alpha)}}\right)\right.\nonumber\\
    &\left.\times\left(\rho_E^{-\frac{\alpha}{2(1+\alpha)}}\frac{1}{2^{(R_1+R_2)}}\sum_{m'_1,m'_2}\rho_{E\lvert X_1\pr{m'_1},X_2\pr{m'_2},X_3\pr{m'_1,m'_2}}\rho_E^{-\frac{\alpha}{2(1+\alpha)}}\right)^\alpha\right]\nonumber\\
    &=\frac{1}{2^{(R_1+R_2)}}\sum_{m_1,m_2}\bbE_C\tra\left[\left(\rho_E^{-\frac{\alpha}{2(1+\alpha)}}\rho_{E\lvert X_1\pr{m_1},X_2\pr{m_2},X_3\pr{m_1,m_2}}\rho_E^{-\frac{\alpha}{2(1+\alpha)}}\right)\right.\nonumber\\
    &\times\left(\rho_E^{-\frac{\alpha}{2(1+\alpha)}}\frac{1}{2^{(R_1+R_2)}}\left(\rho_{E\lvert X_1\pr{m_1},X_2\pr{m_2},X_3\pr{m_1,m_2}}+\sum_{m'_1\ne m_1}\rho_{E\lvert X_1\pr{m'_1},X_2\pr{m_2},X_3\pr{m'_1,m_2}}\right.\right.\nonumber\\
    &\left.\left.\left.+\sum_{m'_2\ne m_2}\rho_{E\lvert X_1\pr{m_1},X_2\pr{m'_2},X_3\pr{m_1,m'_2}}+\sum_{\pr{m'_1,m'_2}\ne\pr{ m_1,m_2}}\rho_{E\lvert X_1\pr{m'_1},X_2\pr{m'_2},X_3\pr{m'_1,m'_2}}\right)\rho_E^{-\frac{\alpha}{2(1+\alpha)}}\right)^\alpha\right]\nonumber\\
    &\mathop\le\limits^{(b)}\frac{1}{2^{(R_1+R_2)}}\sum_{m_1,m_2}\bbE_{\pr{m_1,m_2}}\tra\left[\left(\rho_E^{-\frac{\alpha}{2(1+\alpha)}}\rho_{E\lvert X_1\pr{m_1},X_2\pr{m_2},X_3\pr{m_1,m_2}}\rho_E^{-\frac{\alpha}{2(1+\alpha)}}\right)\right.\nonumber\\
    &\times\left(\rho_E^{-\frac{\alpha}{2(1+\alpha)}}\frac{1}{2^{(R_1+R_2)}}\left(\rho_{E\lvert X_1\pr{m_1},X_2\pr{m_2},X_3\pr{m_1,m_2}}+\sum_{m'_1\ne m_1}\bbE_{\backslash m_1}\rho_{E\lvert X_1\pr{m'_1},X_2\pr{m_2},X_3\pr{m'_1,m_2}}\right.\right.\nonumber\\
    &+\sum_{m'_2\ne m_2}\bbE_{\backslash m_2}\rho_{E\lvert X_1\pr{m_1},X_2\pr{m'_2},X_3\pr{m_1,m'_2}}\nonumber\\
    &\left.\left.\left.+\sum_{\pr{m'_1,m'_2}\ne\pr{ m_1,m_2}}\bbE_{\backslash (m_1,m_2)}\rho_{E\lvert X_1\pr{m'_1},X_2\pr{m'_2},X_3\pr{m'_1,m'_2}}\right)\rho_E^{-\frac{\alpha}{2(1+\alpha)}}\right)^\alpha\right]\nonumber\\
    &=\frac{1}{2^{(R_1+R_2)}}\sum_{m_1,m_2}\bbE_{\pr{m_1,m_2}}\tra\left[\left(\rho_E^{-\frac{\alpha}{2(1+\alpha)}}\rho_{E\lvert X_1\pr{m_1},X_2\pr{m_2},X_3\pr{m_1,m_2}}\rho_E^{-\frac{\alpha}{2(1+\alpha)}}\right)\right.\nonumber\\
    &\times\left(\rho_E^{-\frac{\alpha}{2(1+\alpha)}}\frac{1}{2^{(R_1+R_2)}}\left(\rho_{E\lvert X_1\pr{m_1},X_2\pr{m_2},X_3\pr{m_1,m_2}}+\sum_{m'_1\ne m_1}\rho_{E\lvert X_2\pr{m_2}}\right.\right.\nonumber\\
    &\left.\left.\left.+\sum_{m'_2\ne m_2}\rho_{E\lvert X_1\pr{m_1}}+\sum_{\pr{m'_1,m'_2}\ne\pr{ m_1,m_2}}\rho_E\right)\rho_E^{-\frac{\alpha}{2(1+\alpha)}}\right)^\alpha\right]\nonumber\\
    &\mathop\le\limits^{(c)}\frac{1}{2^{(R_1+R_2)}}\sum_{m_1,m_2}\bbE_{\pr{m_1,m_2}}\tra\left[\left(\rho_E^{-\frac{\alpha}{2(1+\alpha)}}\rho_{E\lvert X_1\pr{m_1},X_2\pr{m_2},X_3\pr{m_1,m_2}}\rho_E^{-\frac{\alpha}{2(1+\alpha)}}\right)\right.\nonumber\\
    &\times\left(\rho_E^{-\frac{\alpha}{2(1+\alpha)}}\frac{1}{2^{(R_1+R_2)}}\left(\rho_{E\lvert X_1\pr{m_1},X_2\pr{m_2},X_3\pr{m_1,m_2}}+2^{R_1}\rho_{E\lvert X_2\pr{m_2}}\right.\right.\nonumber\\
    &\left.\left.\left.+2^{R_2}\rho_{E\lvert X_1\pr{m_1}}+2^{R_1+R_2}\rho_E\right)\rho_E^{-\frac{\alpha}{2(1+\alpha)}}\right)^\alpha\right]\nonumber\\
    &\mathop\le\limits^{(d)}\frac{1}{2^{(R_1+R_2)}}\sum_{m_1,m_2}\bbE_{\pr{m_1,m_2}}\tra\left[\left(\rho_E^{-\frac{\alpha}{2(1+\alpha)}}\rho_{E\lvert X_1\pr{m_1},X_2\pr{m_2},X_3\pr{m_1,m_2}}\rho_E^{-\frac{\alpha}{2(1+\alpha)}}\right)\right.\nonumber\\
    &\times\left(\rho_E^{-\frac{\alpha}{2(1+\alpha)}}\frac{1}{2^{(R_1+R_2)}}\left(\mu_{E,2}\Delta_{E\mid X_1(m_1),X_2(m_2)}\pr{\rho_{E\lvert X_1\pr{m_1},X_2\pr{m_2},X_3\pr{m_1,m_2}}}\right.\right.\nonumber\\
    &\left.\left.\left.+2^{R_1}\mu_{E,1}\Delta_{E\mid X_1(m_1)}\pr{\rho_{E\lvert X_2\pr{m_2}}}+2^{R_2}\mu_E\Delta_E\pr{\rho_{E\lvert X_1\pr{m_1}}}+2^{R_1+R_2}\rho_E\right)\rho_E^{-\frac{\alpha}{2(1+\alpha)}}\right)^\alpha\right]\nonumber\\
    &\mathop\le\limits^{(e)}\frac{1}{2^{(R_1+R_2)}}\sum_{m_1,m_2}\bbE_{\pr{m_1,m_2}}\tra\left[\left(\rho_E^{-\frac{\alpha}{2(1+\alpha)}}\rho_{E\lvert X_1\pr{m_1},X_2\pr{m_2},X_3\pr{m_1,m_2}}\rho_E^{-\frac{\alpha}{2(1+\alpha)}}\right)\right.\nonumber\\
    &\times\left(\rho_E^{-\frac{\alpha^2}{2(1+\alpha)}}\frac{1}{2^{\alpha(R_1+R_2)}}\left(\mu_{E,2}^\alpha\pr{\Delta_{E\mid X_1(m_1),X_2(m_2)}\pr{\rho_{E\lvert X_1\pr{m_1},X_2\pr{m_2},X_3\pr{m_1,m_2}}}}^\alpha\right.\right.\nonumber\\
    &\left.\left.+2^{\alpha R_1}\mu_{E,1}^\alpha\pr{\Delta_{E\mid X_1(m_1)}\pr{\rho_{E\lvert X_2\pr{m_2}}}}^\alpha+2^{\alpha R_2}\mu_E^\alpha\pr{\Delta_E\pr{\rho_{E\lvert X_1\pr{m_1}}}}^\alpha+2^{\alpha(R_1+R_2)}\rho_E^\alpha\Big)\rho_E^{-\frac{\alpha^2}{2(1+\alpha)}}\right)\right]\nonumber\\
    &\mathop=\limits^{(f)}1+\bbE_{X_{[3]}}\tra\left[\frac{\mu_{E,2}^\alpha}{2^{\alpha(R_1+R_2)}}\rho_{E\lvert X_{[3]}}\pr{\Delta_{E|X_1,X_2}\pr{\rho_{E\lvert X_{[3]}}}}^\alpha\rho_E^{-\alpha}\right.\nonumber\\
    &\left.+\frac{\mu_{E,1}^\alpha}{2^{\alpha R_2}}\rho_{E\lvert X_{[3]}}\pr{\Delta_{E\mid X_1}\pr{\rho_{E\lvert X_2}}}^\alpha\rho_E^{-\alpha}+\frac{\mu_E^\alpha}{2^{\alpha R_1}}\rho_{E\lvert X_{[3]}}\pr{\Delta_E\pr{\rho_{E\lvert X_1}}}^\alpha\rho_E^{-\alpha}\right]\nonumber\\
    &\mathop=\limits^{(g)}1+\bbE_{X_{[3]}}\tra\left[\frac{\mu_{E,2}^\alpha}{2^{\alpha(R_1+R_2)}}\rho_{E\lvert X_{[3]}}\Delta_{E\mid X_1,X_2}\pr{\pr{\Delta_{E\mid X_1,X_2}\pr{\rho_{E\lvert X_{[3]}}}}^\alpha\rho_E^{-\alpha}}\right.\nonumber\\
    &\left.+\frac{\mu_{E,1}^\alpha}{2^{\alpha R_2}}\rho_{E\lvert X_{[3]}}\Delta_{E\mid X_1}\pr{\pr{\Delta_{E\mid X_1}\pr{\rho_{E\lvert X_2}}}^\alpha\rho_E^{-\alpha}}+\frac{\mu_E^\alpha}{2^{\alpha R_1}}\rho_{E\lvert X_{[3]}}\Delta_E\pr{\pr{\Delta_E\pr{\rho_{E\lvert X_1}}}^\alpha\rho_E^{-\alpha}}\right]\nonumber\\
    &\mathop=\limits^{(h)}1+\tra\left[\frac{\mu_{E,2}^\alpha}{2^{\alpha(R_1+R_2)}}\bbE_{X_{[3]}}\rho_{E\lvert X_{[3]}}\Delta_{E\mid X_1,X_2}\pr{\pr{\Delta_{E\mid X_1,X_2}\pr{\rho_{E\lvert X_{[3]}}}}^\alpha\rho_E^{-\alpha}}\right.\nonumber\\
    &\left.+\frac{\mu_{E,1}^\alpha}{2^{\alpha R_2}}\bbE_{X_2}\rho_{E\lvert X_2}\Delta_{E\mid X_1}\pr{\pr{\Delta_{E\mid X_1}\pr{\rho_{E\lvert X_2}}}^\alpha\rho_E^{-\alpha}}+\frac{\mu_E^\alpha}{2^{\alpha R_1}}\bbE_{X_1}\rho_{E\lvert X_1}\Delta_E\pr{\pr{\Delta_E\pr{\rho_{E\lvert X_1}}}^\alpha\rho_E^{-\alpha}}\right]\nonumber\\
    &\mathop=\limits^{(i)}1+\tra\left[\frac{\mu_{E,2}^\alpha}{2^{\alpha(R_1+R_2)}}\bbE_{X_{[3]}}\pr{\Delta_{E\mid X_1,X_2}\pr{\rho_{E\lvert X_{[3]}}}}^{1+\alpha}\rho_E^{-\alpha}\right.\nonumber\\
    &\left.+\frac{\mu_{E,1}^\alpha}{2^{\alpha R_2}}\bbE_{X_2}\pr{\Delta_{E\mid X_1}\pr{\rho_{E\lvert X_2}}}^{1+\alpha}\rho_E^{-\alpha}+\frac{\mu_E^\alpha}{2^{\alpha R_1}}\bbE_{X_1}\pr{\Delta_E\pr{\rho_{E\lvert X_1}}}^{1+\alpha}\rho_E^{-\alpha}\right]\nonumber\\
    &\mathop=\limits^{(j)}1+\tra\left[\frac{\mu_{E,2}^\alpha}{2^{\alpha(R_1+R_2)}}\pr{\Delta_{X_{[3]}\otimes E}\pr{\rho_{X_{[3]}E}}}^{1+\alpha}\pr{\rho_{X_{[3]}}\otimes\rho_E}^{-\alpha}\right]\nonumber\\
    &+\tra\left[\frac{\mu_{E,1}^\alpha}{2^{\alpha R_2}}\pr{\Delta_{X_2\otimes E}\pr{\rho_{X_2E}}}^{1+\alpha}\pr{\rho_{X_2}\otimes\rho_E}^{-\alpha}\right]+\tra\left[\frac{\mu_E^\alpha}{2^{\alpha R_1}}\pr{\Delta_{X_1\otimes E}\pr{\rho_{X_1E}}}^{1+\alpha}\pr{\rho_{X_1}\otimes\rho_E}^{-\alpha}\right]\nonumber\\
    &\mathop\le\limits^{(k)}1+\frac{\mu_{E,2}^\alpha}{2^{\alpha(R_1+R_2)}}2^{\alpha\ubar{\D}_{1+\alpha}\pr{\rho_{X_{[3]}E}\big\lVert\rho_{X_{[3]}}\otimes\rho_E}}+\frac{\mu_{E,1}^\alpha}{2^{\alpha R_2}}2^{\alpha\ubar{\D}_{1+\alpha}\pr{\rho_{X_2E}\big\lVert\rho_{X_2}\otimes\rho_E}}\nonumber\\
    &\qquad+\frac{\mu_E^\alpha}{2^{\alpha R_1}}2^{\alpha\ubar{\D}_{1+\alpha}\pr{\rho_{X_1E}\big\lVert\rho_{X_1}\otimes\rho_E}},\label{eq:Resolv_Analysis}
\end{align}where
\begin{itemize}
    \item[$(a)$] follows from the linearity of the trace operation and the expectation;
    \item[$(b)$] follows from the linearity of the expectation and the trace operation and Jensen's inequality;
    \item[$(c)$] follows since $f(x)=x^\alpha$, for $\alpha\in(0\,,1]$, is a matrix monotone function \cite[Section~1.5]{QIT_Hayashi};
    \item[$(d)$] follows from \eqref{eq:Pinching_Prop0};
    \item[$(e)$] follows because the terms inside the second parenthesis of the trace operation commute; additionally, it follows from the inequality $(a+b)^\alpha \le a^\alpha + b^\alpha$ for $\alpha < 1$, and from the property that for $A, B, C \in \calP(\calH)$ with $C \le B$, we have $\tra(AC) \le \tra(AB)$, which holds since $\tra(A(B - C)) \ge 0$ \cite[Lemma~B.5.2]{RennerDissertation};
    \item[$(f)$] follows from the linearity of the trace operation, the symmetry of the codebook construction \ac{wrt} the messages, and since $\Delta_{E\mid X_1,X_2}\pr{\rho_{E\lvert X_{[3]}}}$, $\Delta_{E\mid X_1}\pr{\rho_{E\lvert X_2}}$,  $\Delta_E\pr{\rho_{E\lvert X_1}}$, and $\rho_E$ commute;
    \item[$(g)$] follows since $\Delta_{E\mid X_1X_2}\pr{\rho_{E\lvert X_{[3]}}}$, $\Delta_{E\mid X_1}\pr{\rho_{E\lvert X_2}}$, and $\Delta_E\pr{\rho_{E\lvert X_1}}$ have the same orthonormal basis as $\rho_E$ and therefore 
    \begin{align*}
        \pr{\Delta_{E\mid X_1X_2}\pr{\rho_{E\lvert X_{[3]}}}}^\alpha\rho_E^{-\alpha}&=\Delta_{E\mid X_1X_2}\pr{\pr{\Delta_{E\mid X_1X_2}\pr{\rho_{E\lvert X_{[3]}}}}^\alpha\rho_E^{-\alpha}}\nonumber\\
        \pr{\Delta_{E\mid X_1}\pr{\rho_{E\lvert X_2}}}^\alpha\rho_E^{-\alpha}&=\Delta_{E\mid X_1}\pr{\pr{\Delta_{E\mid X_1}\pr{\rho_{E\lvert X_2}}}^\alpha\rho_E^{-\alpha}}\nonumber\\
        \pr{\Delta_E\pr{\rho_{E\lvert X_1}}}^\alpha\rho_E^{-\alpha}&=\Delta_E\pr{\pr{\Delta_E\pr{\rho_{E\lvert X_1}}}^\alpha\rho_E^{-\alpha}};\nonumber
    \end{align*}
    \item[$(h)$] follows from the linearity of the expectations and the trace operation;
    \item[$(i)$] follows from \eqref{eq:Pinching_Switching};
    \item[$(j)$] follows from the definition of pinching operations $\Delta_E$ , $\Delta_{E\lvert X_1}$, and $\Delta_{E\lvert X_1,X_2}$  and since the involved states are classical-quantum states;
    \item[$(k)$] follows from the definition of $\ubar{\D}_{1+\alpha}\big(\cdot\lVert\cdot\big)$ and since for two quantum states $\rho,\sigma\in\calD(\calH)$ and quantum operation $\Upsilon(\cdot):\calL(A)\to\calL(B)$, we have \cite{Branum96,Frank13},
    \begin{align}
        \ubar{\D}_{1+\alpha}\big(\Upsilon(\rho)\lVert\Upsilon(\sigma)\big)\le\ubar{\D}_{1+\alpha}\big(\rho\lVert\sigma\big).\label{eq:Monotonicity_SRenyi}
    \end{align}
\end{itemize}
Now, substituting \eqref{eq:Resolv_Analysis} into \eqref{eq:Jensen_Res} and using the inequality $\log_2(1+x) \le \frac{x}{\ln 2}$ completes the proof of Lemma~\ref{lemma:Resolvability}.

\section{Proof of Theorem~\ref{thm:Achievable_Asymp}}
\label{proof:thm:Achievable_Asymp}
We begin by bounding $\mu_B$ and $\mu_{B,3}$, as defined in \eqref{eq:gs}, under the assumption that the transmitters use the channel $n$ times in an \ac{iid} manner. That is, we consider the scenario in which there are $n$ \ac{iid} copies of the states $\rho_{X_{[3]}B}$, $\rho_{\pr{X_2,X_3}-X_1-B}$, and $\rho_{\pr{X_1,X_3}-X_2-B}$. The following lemmas are essential in our proof.
\begin{lemma}
\label{eq:Lemma_Pincing_Eigen_Value_Bounds}
Let $ \Delta_E $, $ \Delta_B $, and $ \Delta_1 $ be the pinching maps defined in Section~\ref{sec:Pinching}, and let $ \mu_E $, $ \mu_B $, and $ \mu_{B,3} $ be as defined in \eqref{eq:gs}. Denote by $d_{X_1}$, $d_{X_2}$, $d_{X_3}$, and $d_B$ the dimensions of the Hilbert spaces $\calH_{X_1}$, $\calH_{X_2}$, $\calH_{X_3}$, and $\calH_B$, respectively. Then the following bounds hold:
\begin{subequations}\label{eq:Upper_Bounds_g1_g2}
\begin{align}
    \mu_E &\le (n+1)^{d_E - 1},\\
    \mu_{E,1} &\le (n+1)^{\frac{d_{X_1}}{2}(d_E + 2)(d_E - 1)},\\
    \mu_{E,2} &\le (n+1)^{\frac{d_{X_1}d_{X_2}}{2}(d_E + 2)(d_E - 1)},\\
    \mu_B &\le (n+1)^{d_B - 1},\\
    \mu_{B,3} &\le (n+1)^{\frac{d_{X_1}d_{X_2}d_{X_3}}{2}\pr{d_B + 2}\pr{d_B - 1}}.
\end{align}
\end{subequations}
\end{lemma}
The proof of Lemma~\ref{eq:Lemma_Pincing_Eigen_Value_Bounds} follows similar steps to those in the proof of \cite[Proposition~12]{Salek25} and is omitted for brevity.
\begin{lemma}
\label{eq:Ryini_Conditional_MI}
For each $\alpha\in\pr{\frac{-1}{2}:0}\cup\pr{0:\infty}$, we have
\begin{align}
    &\ubar{\I}_{1-\alpha}\pr{X_1,X_3;B\lvert X_2}_{\rho_{X_{[3]}B}^{\otimes n}\lvert\rho_{X_{[3]}}^{\otimes n}}=n\ubar{\I}_{1-\alpha}\pr{X_1,X_3;B\lvert X_2}_{\rho_{X_{[3]}B}\lvert\rho_{X_{[3]}}},
    \label{eq:Riyni_Entropy_1}\\
    &\ubar{\I}_{1-\alpha}\pr{X_2,X_3;B\lvert X_1}_{\rho_{X_{[3]}B}^{\otimes n}\lvert\rho_{X_{[3]}}^{\otimes n}}=n\ubar{\I}_{1-\alpha}\pr{X_2,X_3;B\lvert X_1}_{\rho_{X_{[3]}B}\lvert\rho_{X_{[3]}}}.
    \label{eq:Riyni_Entropy_2}
\end{align}
\end{lemma}The proof of Lemma~\ref{eq:Ryini_Conditional_MI} is similar to the proof of \cite[Lemma~4]{AshnuHayashi2020} and is omitted for brevity. From Theorem~\ref{thm:Achievable_One_Shot} when the transmitters use the channel $\calN_{X_{[T]}\to BE}$ $n$ times independently, there exists a code such that
\begin{subequations}\label{eq:Pe_Covertness_2}
\begin{align}
    \bbP\left\{\hat{M}\ne M\right\}&\le6\mu_{B,3}^\alpha\left(2^{\alpha\pr{R_1+R_2-\ubar{\D}_{1-\alpha}\pr{\rho_{X_{[3]}B}^{\otimes n}\lVert\rho_{X_{[3]}}^{\otimes n}\otimes\rho_B^{\otimes n}}}}+2^{\alpha\pr{ R_1-\ubar{\I}_{1-\alpha}\pr{X_1^n,X_3^n;B^n\lvert X_2^n}}}\right.\nonumber\\
    &\quad\left.+2^{\alpha\pr{ R_2-\ubar{\I}_{1-\alpha}\pr{X_2^n,X_3^n;B^n\lvert X_1^n}}}\right),\label{eq:Pe_Thm_2}\\
    \left\lVert\tau_E- \rho_0\right\rVert_1&\le\frac{2}{\sqrt{\alpha}}\left(\pr{\mu_{E,2}}^{\frac{\alpha}{2}}2^{\frac{\alpha}{2}\pr{-(R_1+R_2)+\ubar{\D}_{1+\alpha}\pr{\rho_{X_{[3]}E}^{\otimes n}\big\lVert\rho_{X_{[3]}}^{\otimes n}\otimes\rho_E^{\otimes n}}}}\right.\nonumber\\
    &\qquad+\pr{\mu_{E,1}}^{\frac{\alpha}{2}}2^{\frac{\alpha}{2}\pr{-R_1+\ubar{\D}_{1+\alpha}\pr{\rho_{X_1E}^{\otimes n}\big\lVert\rho_{X_1}^{\otimes n}\otimes\rho_E^{\otimes n}}}}\nonumber\\
    &\qquad\left.+\pr{\mu_E}^{\frac{\alpha}{2}}2^{\frac{\alpha}{2}\pr{-R_2+\ubar{\D}_{1+\alpha}\pr{\rho_{X_2E}^{\otimes n}\big\lVert\rho_{X_2}^{\otimes n}\otimes\rho_E^{\otimes n}}}}\right).\label{eq:Covertnes_Thm_2}
    \end{align}
\end{subequations}
Now using Lemma~\ref{eq:Lemma_Pincing_Eigen_Value_Bounds} and Lemma~\ref{eq:Ryini_Conditional_MI} and \eqref{eq:Pe_Covertness_2}, when $\alpha\to0$ and $n$ grow to infinity, there exists a sequence of codes such that \eqref{eq:Pe_Asymp} and \eqref{eq:CC_CSK_Metric_Asymp} are satisfied if, we have
\begin{subequations}\label{eq:Initial_Bound}
\begin{align}
    R_1&<I(X_1,X_3;B\lvert X_2),\label{eq:Initial_Bound_Rel_1}\\
    R_2&<I(X_2,X_3;B\lvert X_1),\label{eq:Initial_Bound_Rel_2}\\
    R_1+R_2&<I(X_1,X_2,X_3;B),\label{eq:Initial_Bound_Rel_3}\\
    R_1&>I(X_1,X_3;E\lvert X_2),\label{eq:Initial_Bound_Res_1}\\
    R_2&>I(X_2,X_3;E\lvert X_1),\label{eq:Initial_Bound_Res_2}\\
    R_1+R_2&>I(X_1,X_2,X_3;E).\label{eq:Initial_Bound_Res_3}
\end{align}
\end{subequations}
Combining the rate constraints in \eqref{eq:Initial_Bound} leads to the achievable rate region in Theorem~\ref{thm:Achievable_Asymp}.

\bibliographystyle{IEEEtran}
\bibliography{IEEEabrv,bibfile}

\end{document}

%% file: CommandsAndMacros.tex
\newcommand{\calC}{\mathcal{C}}

\newcommand{\calD}{\mathcal{D}}
\newcommand{\bbD}{\mathbb{D}}

\newcommand{\bbE}{\mathbb{E}}

\newcommand{\calF}{\mathcal{F}}

\newcommand{\calH}{\mathcal{H}}

\newcommand{\calI}{\mathcal{I}}
\newcommand{\bbI}{\mathbb{I}}

\newcommand{\calJ}{\mathcal{J}}

\newcommand{\calL}{\mathcal{L}}

\newcommand{\calM}{\mathcal{M}}

\newcommand{\calN}{\mathcal{N}}
\newcommand{\bbN}{\mathbb{N}}

\newcommand{\calP}{\mathcal{P}}
\newcommand{\bbP}{\mathbb{P}}

\newcommand{\calR}{\mathcal{R}}
\newcommand{\bbR}{\mathbb{R}}

\newcommand{\calS}{\mathcal{S}}

\newcommand{\calX}{\mathcal{X}}

\newcommand\ubar[1]{\stackunder[1.1pt]{$#1$}{\rule{1.2ex}{.08ex}}}

\newcommand{\brk}[1]{\ensuremath{\big[{#1}\big]}}

\DeclareMathOperator{\tra}{Tr}
\DeclareMathOperator{\D}{D}
\DeclareMathOperator{\I}{I}

\DeclareMathOperator{\F}{F}
\DeclareMathOperator{\Pu}{P}

\newcommand{\den}[2]{\ensuremath{\ket{#1}{\hspace{-1.6mm}}\bra{#2}}}

\newtheorem{theorem}{Theorem}
\newtheorem{corollary}{Corollary}
\newtheorem{definition}{Definition}

\newtheorem{lemma}{Lemma}[]

\newcommand{\indic}[1]{\ensuremath{\mathds{1}}}
\newcommand{\card}[1]{\ensuremath{\left\lvert{#1}\right\rvert}}   

\newcommand{\sbra}[2]{\ensuremath{\left[{#1}{\,:\,}{#2}\right]}}%
\newcommand{\sbr}[1]{\ensuremath{\left[{#1}\right]}}%
\newcommand{\pr}[1]{\ensuremath{\left({#1}\right)}}%
\newcommand{\br}[1]{\ensuremath{\left\{{#1}\right\}}}%



%% file: Acronyms.tex
\acrodef{ACDIS}[ACDIS]{Adaptive Communication Decision and Information Systems}
\acrodef{AEP}{Asymptotic Equipartition Property}
\acrodef{AoA}{Angle of Arrival}
\acrodef{AWGN}{Additive White Gaussian Noise}
\acrodef{AVC}[AVC]{Arbitrarily Varying Channel}
\acrodef{BER}{Bit-Error-Rate}
\acrodef{BEC}{Binary Erasure Channel}
\acrodef{BPSK}{Binary Phase-Shift Keying}
\acrodef{BSC}{Binary Symmetric Channel}
\acrodef{RV}{Random Variable}
\acrodefplural{RV}{Random Variables}
\acrodef{MP}{Multi-Parent}
\acrodef{JTE}{Joint Typical Encoder}
\acrodef{BICM}[BICM]{Bit-Interleaved Coded-Modulation}
\acrodef{CDF}[CDF]{Cumulative Distribution Function}
\acrodef{CGF}[CGF]{Cumulant Generating Function}
\acrodef{CLT}[CLT]{Central Limit Theorem}
\acrodef{CSI}[CSI]{Channel State Information}
\acrodef{DMC}[DMC]{Discrete Memoryless Channel}
\acrodef{DMS}[DMS]{Discrete Memoryless Source}
\acrodef{ERM}[ERM]{Empirical Risk Minimization}
\acrodef{FER}[FER]{Frame Error Rate}
\acrodef{ICA}[ICA]{Independent Component Analysis}
\acrodef{iid}[i.i.d.]{independent and identically distributed}
\acrodef{IoT}[IoT]{Internet of Things}
\acrodef{KKT}[KKT]{Karush-Kuhn Tucker}
\acrodef{LASSO}[LASSO]{Least Absolute Shrinkage and Selection Operator}
\acrodef{LPD}[LPD]{Low Probability of Detection}
\acrodef{LDPC}[LDPC]{Low-Density Parity-Check}
\acrodef{LLMS}[LLMS]{Linear Least Mean Square}
\acrodef{LMS}[LMS]{Least Mean Square}
\acrodef{MAC}[MAC]{multiple-access channel}
\acrodef{QMAC}[QMAC]{Quantum Multiple-Access Channel}
\acrodef{MGF}[MGF]{Moment Generating Function}
\acrodef{MLC}[MLC]{Multi-Level Coding}
\acrodef{MLE}[MLE]{Maximum Likelihood Estimate}
\acrodef{MIMO}[MIMO]{Multiple-Input Multiple-Output}
\acrodef{MISO}{Multiple-Input Single-Output}
\acrodef{MSD}[MSD]{Multi-Stage Decoding}
\acrodef{MMSE}[MMSE]{Minimum Mean-Square Error}
\acrodef{PAC}[PAC]{Probably Approximately Correct}
\acrodef{PCA}[PCA]{Principal Component Analysis}
\acrodef{PDF}[PDF]{Probability Density Function}
\acrodefplural{PDF}{Probability Density Functions}
\acrodef{PMF}[PMF]{Probability Mass Function}
\acrodefplural{PMF}{Probability Mass Functions}
\acrodef{PPM}[PPM]{Pulse Position Modulation}
\acrodef{PSD}{Power Spectral Density}
\acrodef{PSK}{Phase Shift Keying}
\acrodef{QKD}{Quantum Key Distribution}
\acrodef{ROC}{Receiver Operating Characteristic}
\acrodef{CVQKD}{Continuous-Variable \ac{QKD}}
\acrodef{QPSK}{Quadrature Phase-Shift Keying}
\acrodef{RV}{random variable}
\acrodef{SIMO}{Single-Input Multiple-Output}
\acrodef{SNR}{Signal-to-Noise Ratio}
\acrodef{SVM}[SVM]{Support Vector Machine}
\acrodef{POVM}{Positive Operator-Valued Measure}
\acrodefplural{POVM}{Positive Operator-Valued Measures}
\acrodef{wrt}[w.r.t.]{with respect to}
\acrodef{WSS}{Wide Sense Stationary}
\acrodef{RHS}{Right Hand Side}
\acrodef{LHS}{Left Hand Side}
\acrodef{CPTP}{Completely Positive and Trace Preserving}
\acrodef{ADSI}[ADSI]{Action-Dependent State Information}